\documentstyle[11pt,aasms4]{article}
\def\etal{{\it et al.\ }}
\def\eg{{\it e.g., \,}}
\def\ie{{\it i.e., \,}}

\def\putplot#1#2#3#4#5#6#7{\begin{centering} \leavevmode
\vbox to#2{\rule{0pt}{#2}}
\includegraphics{#1}
\end{centering}}
\begin{document}
\title{The Nature of a Dusty Ring in Virgo}

\author{Noah Brosch \& Elchanan Almoznino} 
\affil{The Wise Observatory and 
the School of Physics and Astronomy,
Raymond and Beverly Sackler Faculty of Exact Sciences,
Tel Aviv University, Tel Aviv 69978, Israel}

\author{Bogdan Wszolek\altaffilmark{1} \& Konrad Rudnicki}
\affil{Jagiellonian University Astronomical Observatory, ul. Orla 171, PL-30-244 Krakow, Poland}
\altaffiltext{1}{Also at the  Pedagogical University, ul. Armii Krajowej, PL-42-207 Czestochowa, 
Poland.}

%\maketitle

\begin{abstract}
We detected a ring-like distribution of far-infrared emission in the direction of the 
center of the Virgo cluster. We studied this feature in the FIR, radio, and optical domains,
and deduced that the dust within the feature reddens the galaxies in the direction of the Virgo cluster 
but does not affect stars within the Milky Way. This is likely to be a dusty feature in the 
foreground of the Virgo cluster, presumably in the galactic halo. The HI 
distribution follows the morphology of the FIR emission and shows peculiar kinematic
behavior. We propose that a highly supersonic past collision between an HI cloud
and the Galactic HI formed a shock that heated the interface gas to soft
X-ray temperatures.  HI remnants from the
projectile and from the shocked Galactic HI rain down onto the disk as intermediate
velocity gas.

Our finding emphasizes that extragalactic astronomy must consider the possibility of
extinction by dust at high Galactic latitude and far from the Galactic plane, which
may show structure on one-degree and smaller scales.
This is particularly important for studies of the Virgo cluster, for
example in the determination of the Hubble constant from Cepheids in cluster galaxies.

\end{abstract}


\section{Introduction}
The nature of non-luminous matter that is not part of
detected and catalogued galaxies remains unsolved by modern 
astrophysics. As mentioned in
a recent thesis, low surface brightness (LSB) objects may prove to be the ``icebergs'' of the 
extragalactic world (de Blok 1997). Some searches for non-luminous matter have been successful,
\eg the detection of a giant HI ring around the small group of galaxies in Leo centered on M96
    (Schneider \etal 1983), extended HI emission in the M81 group (Lo \&
    Sargent 1979), HI companions to dwarf galaxies (for $\sim$25\% of the
    cases: Taylor \etal 1996), and a large neutral hydrogen cloud in the
    southern outskirts of the Virgo cluster (HI 1225+01: Giovanelli \&
    Haynes 1989).

Along with HI clouds, a few large LSB galaxies have been identified: Malin-1
(Bothun \etal 1987), F568-6 (Bothun \etal 1990), and
    1226+0105 (Sprayberry \etal 1993). Their typical star formation rates are
    $\sim$0.1 M$_{\odot}$/yr and the metallicities are  $\sim$1/3 solar. 
The HI rotation curves, measured by de Block \etal (1997) and by
    Pickering \etal (1997), indicate that their gaseous component is
    dynamically significant at all radii and that the galaxies are fully dark-matter
    dominated; their detected baryonic component is less than 4\% of
   the total mass. This last conclusion is valid at least as long as we do not accept any of the more
exotic theories of gravitation. The LSB galaxies lack bulges, bars, and nuclear activity, as well as
    CO or IR emission (\ie have no molecules or dust).

There have also been a few intriguing reports of presumably 
intergalactic dust clouds. A cloud
with 0.5-1.2 mag of extinction was identified in Microscopium by Hoffmeister (1962). Three other
similar objects were listed by Rudnicki (1986); they extinguish background objects by 
0.57 to 1.2 mag. In all reports the main point of contention was the
actual distance to the cloud, which could put it in extragalactic space but could also locate it
in the halo of the Milky Way (MW). Sometimes, the argument for an extragalactic location was
based on a comparison of the properties of objects whose distance could be estimated
and which were located behind the cloud with those of similar objects clearly
not within the cloud limits (\eg RR Lyrae stars; Murawski 1983).

The extragalactic nature is only fairly confidently
established for the Abadi-Edmunds cloud at $\sim$3 Mpc (Abadi \& Edmunds 1978). 
HI 21 cm line emission was detected from this
object, whereas in other cases it was not. However, in other cases far-infrared (FIR) 
emission was detected and
 could be identified (on morphological and positional criteria) with the obscuring
clouds. FIR and HI emission were clearly detected in the case of the Okroy cloud (Wszolek 
\etal 1988a, 1989). FIR  emission
was only marginally detected from the Rudnicki-Baranowska cloud (Wszolek \etal 1988b). 
This indicates that the physical
conditions in this kind of objects are far from being uniform. More such examples must be 
identified and their properties examined.

It is possible that the phenomenon of intergalactic hydrogen clouds could be related 
to the high-velocity
cloud (HVC) complexes. These are HI structures whose radial velocities deviate by 
several 100 km s$^{-1}$
from the conventional galactic rotation. A recent review of HVCs is by Wakker \& van Woerden (1997).
Their Table 2 lists a few cloud complexes at distances $\geq$25 kpc; some of these may not
belong at all to the MW.
IRAS searches for FIR emission of HVC were negative (\eg Wakker \&
Boulanger 1986), indicating that either the HVCs are dust-free or that their dust
grains are much cooler than could be detected with IRAS. In this context we also mention 
the proposition by Blitz \etal (1999) that the HVCs make up the missing mass by
being essentially dark halos with low velocity dispersions.

We report here results from a study of a diffuse ring-like FIR feature at high galactic latitude, which
we interpret as ``local'', \ie not extragalactic, despite first indications to the contrary.
The region toward which this feature is located is the center of the Virgo cluster of
galaxies. This part
of the sky has been studied in exquisite detail, yet new studies always detect interesting
features. For example, Katsiyannis \etal (1998) produced a very deep image of the central
regions of the cluster from a combination of 13 deep Kodak TechPan films obtained with the
UK Schmidt telescope. The image shows large variations in the brightness of the
intra-cluster medium, with the brightest regions north of the cluster center. M87 is  
fairly central in the region of enhanced brightness, close to the upper left corner of the
``very high contrast image'' in their Fig. 6.
Previous deep imaging of the central VC region (\eg Weil \etal 1997) revealed a diffuse 
extension of (presumably stellar) material extending $\sim$100 kpc to the SE of M87. Intergalactic
red giant stars were apparently discovered near M87 by Ferguson \etal (1998). It is therefore
relevant to search for, and to try and explain, any extended feature one may detect in the
direction of the center of the cluster. In this context, we mention
 the study of Haikala \etal (1995) who examined the UV emission detected
in the direction of a dust globule close to the North Galactic Pole, slightly north
of the Virgo cluster (VC).

Any material that could produce 
extinction needs to be accounted for. To the best of our knowledge, nobody attempted to study
the obscuration and FIR emission by ISM or IGM in the direction of a rich, nearby cluster
of galaxies.
This is particularly important for the VC, which serves as one of the key stones in 
the distance ladder leading up to
the determination of the Hubble constant (\eg van den Bergh 1996). The HST Key
Project on the Extragalactic Distance Scale, where the required accuracy of the
determination of H$_0$ is 10\%, could be affected significantly by unaccounted extinction.
Until now, seven galaxies within 10$^{\circ}$ of the Virgo center have been observed
for Cepheids in this context (Macri \etal 1999).

The plan of the paper is as follows: we first describe the FIR observations, which revealed the
feature, and present confirmatory evidence of its reality. We then attempt to derive
additional properties of the feature, which has an approximate ring shape, using data in the
optical and radio domains. We show that the dust in the feature does not seem to affect the
stars in the Milky Way but that it apparently reddens galaxies in the VC and beyond. The 
full data set is discussed in the last section of the paper,
in which we also derive some properties of the dust grains in the feature.

\section{Observational data}

\subsection{COBE/DIRBE}

Far infrared (FIR) observations from the COBE satellite, specifically with the DIRBE instrument,
reveal non-uniform FIR emission from the center of the VC.
The DIRBE instrument mapped the entire sky at ten wavelength bands from 1.25 to
240$\mu$m and operated from November 1989 to December 1993 (cryo-cooling  
was available only for ten months, restricting the availability of the FIR channels). 
An important feature of DIRBE was that the measurements 
were performed against an internal calibrator source, with proper accounting for
instrumental offsets and interplanetary FIR emission. For the present analysis we used the
Annual Average Sky Maps (AASM: Hauser \etal 1997), which provide a single, ten-month averaged
intensity value per pixel in each of the DIRBE bands. Note that the zodiacal light contribution
was not subtracted from the DIRBE counts. This is because we do not estimate the zodiacal
contribution to the FIR bands to be significant or
to show features on the angular scales relevant here.

We conducted a number of studies of galaxies in the Virgo cluster in which we studied
various photometric indices for entire objects as well as for localized regions in
each galaxy (\eg Almoznino \& Brosch 1998, Heller \etal 1998). The possibility
that these programs could be affected by foreground dust imposed our selection of
the Virgo Cluster as the
initial target for combined FIR and other spectral band interpretations.

We detected a ring-like structure of FIR emission in COBE/DIRBE maps of the VC, which is
centered approximately on M87. The ring is approximately centered on (1950) 12$^h$31$^m$; 
+13$^{\circ}$ (l=285$^{\circ}$.8, b=75$^{\circ}$; J2000) and its diameter is $\sim4^{\circ}$. The
width of the FIR emission in the rim of the ring is $\sim1^{\circ}$.
The detection was made originally on the COBE/DIRBE
maps, but the existence of the feature was established also on IRAS maps (see below).
The M87 galaxy (l$\approx282^{\circ}$.5, b$\approx+74^{\circ}$.4) is normally
taken as the center of the Virgo Cluster (VC) and one could imagine scenarios by which some sort
of FIR-emitting matter could be distributed around it. For this reason, we decided to follow
the FIR detection of the feature, which we call here ``the Virgo Ring'' (VR), and investigate
it further. 

  The detection was made on the AASM, which have noise levels of 3 10$^{-3}$ MJy sr$^{-1}$
at 100$\mu$m, 0.6 MJy sr$^{-1}$ at 140$\mu$m, and 0.3 MJy sr$^{-1}$ at 240$\mu$m (Kashlinsky
1999).
 The ring is visible even by superficial inspection of these COBE/DIRBE 
 gray scale maps. No traces of the ring can be seen on 60$\mu$m or shorter wavelength maps.
 To obtain detailed insight into the structure of the VR we produced isophotal maps 
at $\lambda$=100 and 240$\mu$m using the
 original 0$^{\circ}$.3 square pixels, which are shown as isophote plots in Figure 1. 
The 100$\mu$m map shows a region of depressed FIR flux where
F$_{100}\approx$8.2 MJy sr$^{-1}$. This is surrounded by regions of enhanced FIR emission,
which reach F$_{100}\approx$10 MJy sr$^{-1}$. The 240$\mu$m map indicates that the region of
reduced FIR emission has F$_{240}\approx$3.7 MJy sr$^{-1}$ while the surrounding regions
have F$_{240}\approx$5 MJy sr$^{-1}$. It is clear that (a) the DIRBE data indicate a 
region of low FIR emission surrounded by enhanced emission, and (b) the feature is
real, because it appears on more than one DIRBE map. The lowest values of the FIR flux 
originate presumably from the zodiacal light that was not subtracted from the AASMs
and from the cosmic FIR background. As both these components are much smoother than
the feature we describe here, there is no need to model them in detail.

\subsection{IRAS}

The peculiar FIR features detected by COBE/DIRBE are confirmed by IRAS measurements.
The IRAS mission mapped the sky in four wavelength bands from January 1983 to November 1983.
The primary goal of the IRAS survey was the detection of point sources, but a catalog of
extended sources has also been produced, as well as sky brightness images in each
of the four bands with 2' pixels and 4'-6' resolution (Beichman \etal 1988).

   IRAS 60 and 100$\mu$m Extended Emission Data in the 16$^{\circ}.5\times16^{\circ}.5$ 
square fields no. 83 and 84
  were used to confirm the existence of the ring and to exclude the possibility of
instrumental artefacts produced by the COBE/DIRBE instrument. 
We created maps at these two spectral bands  with a 4'$\times$4' beam. 
The VR is clearly visible 
on the  100$\mu$m map shown in Fig. 1. A similar 100$\mu$m  map based 
on IRAS observations, and where this feature is also visible, was reproduced already
by Leggett \etal (1987) as their Plate 2. The enhanced  IRAS resolution 
relative to COBE/DIRBE allows a good
 morphological evaluation of the FIR feature. In addition to the north-westerly extension
of the FIR emission, along the IRAS scan direction, one sees an arc-like distribution of
emission, which could be interpreted as forming an elliptical ring. Note that the
feature is visible only on the 100$\mu$m map (shown in Figure 1) and  is not  seen 
on the 60$\mu$m map,
or on those at even shorter wavelengths (not shown here).

Although the low resolution COBE/DIRBE maps seem to indicate that the FIR emission is arranged
in a ring, with low FIR at the center and high emission on its perimeter, the higher 
resolution IRAS maps show that this is not the case. The FIR emission
is distributed in an open configuration, with a region of low emission centered on
$\sim12^h30^m$, +13$^{\circ}$.2.  The FIR emission could best be described as
a fork, or a two-arc shape limited to $\alpha$=185$^{\circ}-189^{\circ}$. The
eastern side of the feature shows a small region of enhanced FIR emission centered on
$\alpha$=185$^{\circ}$ and  $\delta$=13$^{\circ}$.5 that stands out over its surroundings and to which
we refer as the ``main blob'' (MB).

\subsection{Optical information: stars}

The dust revealed by the FIR observations may (a) extinguish and
(b) redden stars behind it. The first effect is a consequence of the ``total
extinction'' property, whereas the second is the result of ``wavelength-selective
extinction''. The relative importance of the two effects is linked through the
parameter $R=\frac{A_V}{E(B-V)}$, which is determined to first order by the size of the dust 
grains.

We tested two assumptions, one of extinction within the Milky Way (MW) that would affect some of
the stars but not others, and a second that the VR is extragalactic and is located between the 
MW and the VC. In the second case it would affect the VC galaxies, but none of the MW stars.

For testing the possibility that the dust is ``local'' one requires a  large number 
of stars with magnitudes and colors. These were extracted from the USNO-A2.0 catalog, which
includes blue and red magnitudes for each star. The
USNO-A2.0 catalog contains $>$5 10$^8$ objects ($\sim$12,750 per square degree) and is based 
on scans of the Palomar Sky 
Survey (PSS) plates produced with the Precision Measuring Machine (PMM). The catalog is an
improvement over the version 1.0 both in astrometric accuracy and in photometric precision.
The photometric
accuracy is probably not better than $\sim$0.15 mag, but the depth of the catalog is considerable,
as it reaches 20-22 mag (color-dependent). It can, therefore, serve as a source of 
stellar objects with which one can test the assumption of foreground extinction.

We extracted objects in a number of $1^{\circ}\times1^{\circ}$ regions from the USNO-A2.0 
catalog. The extraction locations are listed in Table 1 and correspond to some FIR-bright
regions (where we expect a higher density of extinguishing dust) or to some FIR-faint
regions (which should be $\sim$transparent). We produced Wolf diagrams for each location,
and show these in Figure 2. The Wolf diagrams plot the cumulative distribution of
stellar magnitudes against magnitude, and the signature of total extinction in such
a plot is a step-like deviation, to fainter magnitudes of the cumulative star counts, 
from the pattern set by 
the brighter (and closer, on average) stars. The diagrams do not show such a 
step-like trend for
regions in the direction of stronger FIR emission when compared with the behavior of the
cumulative distribution in regions with lower FIR emission.

It is also possible to compare the measured behavior of the cumulative star counts 
with that ``predicted'' in absence of localized 
extinction effects by using a model for the stellar distribution in 
the Galaxy for the same Milky Way locations as sampled here. A very successful 
and intuitively simple stellar
distribution model was produced by Bahcall \& Soneira (1984) 
and is available on-line\footnote{http://www.sns.ias.edu/$\sim$jnb/Html/galaxy.html}. 
We calculated predicted star counts for the
locations of the extracted data from the USNO-A2.0 catalog using the
version of the model retrieved in December 1998. The locations are listed in Table
1. We compared the predicted cumulative star counts with the actual star counts. 
The comparisons are shown in Figure 3
and  show no significant deviations from the predicted behavior.

The exercises shown in Figs. 2 and 3 indicate that the stellar distributions are not
influenced by the material producing the FIR emission. The conclusion is, therefore,
that this material is either extremely nearby, so that all the stars are affected
in the same manner, or that it is very distant, beyond the more distant stars listed in the 
USNO-A2.0 catalog.

\subsection{Optical information: galaxies}

If the dust observed in the FIR does not affect stars in our galaxy, it may be located
far from the MW and could affect only objects seen behind it.
Testing the assumption of a dust cloud distant from the MW  requires a sample of background
objects with relatively high surface density, as well as brightness
and color information. In the Virgo region, the ``standard'' extragalactic catalog has been 
for a number of years the
Binggeli \etal (1985) Virgo Cluster Catalog (VCC). The VCC covers $\sim$140 square degrees
and contains 2096 galaxies. The surface density of galaxies is, therefore, $\sim$15 
galaxies/square degree, on average. While this may appear sufficient, the photometry is
not adequate because the galaxy magnitudes  
in the VCC are eye estimates and may have significant deviations. In addition,
no colors are available for most VCC galaxies. We decided therefore to rely
on a more recent galaxy compilation, which reaches deeper in brightness and is thus
denser than the VCC, has better
photometry, and contains color information for the objects.

Currie \& Young (1998, hereafter VPC) produced an extensive three-color photometric catalog 
of galaxies in the
central regions of the VC. The catalog is based on COSMOS scans of one U plate, two
B$_J$ plates, and one R$_C$ plate, all obtained with the UK Schmidt telescope. The
plates were photometrically calibrated and objects were extracted automatically, with
stars and galaxies separated by an automatic algorithm. The VPC provides an impartial
 survey of galaxies in the region of interest for the present study, which reaches to 
B$_J\approx$19 mag, thus it is somewhat shallower but comparable in depth with the stars 
from the USNO-A2.0 catalog.
The area covered by the VPC covers 23 square degrees and is centered on (1950) 12:26 +13:08.
The average galaxy surface density is therefore 49 galaxies/square degree, considerably more than
that of the VCC.

We attempted to detect total extinction effects on the VPC galaxies by limiting
the analysis to regions with high FIR emission and comparing these with similar
analyses in the direction of regions with lower FIR emission. Four parallerogram-shaped 
fields were selected,  marked
A, B, C, and D on Figure 4. Fields C and D are used to determine the nature
of the galaxy population in the general region of the VR. Field A  could also be used 
for this purpose, but we caution that
a background cluster of galaxies (Abell 1552 at z$\approx$0.084) is located in
this field and thus region A may not be representative. This galaxy cluster is presumably 
part of a background 
sheet-like complex, which includes also Abell 1526 at a very similar redshift.
Field A was selected to offer insight on how the presence of background galaxies
disturbs the results. The enhancements of the galaxy background
may distort the Wolf diagrams of galaxies (Figure 5), and indeed some FIR enhancements could be
in the direction of these galaxy clusters. However, searches for dust in clusters of galaxies
have, so far, been negative (Maoz 1995). Thus, we may tentatively discount the FIR enhancements
in the direction of background clusters of galaxies as chance superpositions. Field B is considered  
not to be affected by absorption/extinction and is in the direction of the low
FIR emission of the VR.

Attempts to detect the presence of dust as a ``total extinction'' effect, which
modifies the cumulative galaxy counts between the different regions,
were not successful. The differences were not significant and indicate that if
dust is present, it may cause at most a small amount of total extinction: A$_B\leq$0.5 mag.
We therefore checked for the presence of color-dependent extinction by studying the
distribution of the (U--R$_C$) color index in one square degree areas over the
central part of the Virgo cluster. The data used for this test, and an extensive
description of the method and results, are given in the Appendix to this paper. Here
we emphasize that the results show that the galaxies in the direction of the Virgo
Ring (VR)
part with the lowest FIR emission appear slightly bluer than those in the direction of the
two regions with higher FIR emission. The difference is significant to $\geq$95\%.
Interpreted as dust extinction, this difference in average (U--R$_C$) color index 
indicates a possible  wavelength-dependent extinction of 
$\Delta$(U--R$_C$)$\simeq$0.3 mag between areas with high FIR emission and areas with less dust, 
\ie a total extinction A$_V\simeq$0.33 for a typical Milky Way extinction law, although 
this was not checked here.

\subsection{Radio information}

Here we show that X-ray observations of the region indicate a two-component makeup for
the hot gas, and that the morphology and kinematics of the HI are peculiar.
B\"{o}hringer \etal (1995) mapped the X-ray emission from the immediate vicinity of M87; 
this is the region of
interest of the present study. Their findings show the presence of thermal X-ray
emission from cooler gas than the intracluster medium. A ROSAT map of the general region, larger than
the one analyzed in the 1995 paper, was presented by B\"{o}hringer \etal (1994) and shows a ridge of 
X-ray emission which approximately coincides with the FIR emission ridge to the west of M87. 
They mention, in particular, the sharp drop in X-ray intensity on the western side of M87. 
B\"{o}hringer \etal (1994) subtracted a model distribution of X-rays from M87 from the ROSAT map
and derived a residual map (their Fig. 2) which shows the background cluster A1552 at 12:30+11:30 and
a long filament, which is elongated $\sim$north-south at $\alpha\approx$12$^h$30$^m$ and from 
$\delta\approx$+15 to +6. This filament curves around M87 on its westerly side and seems to follow the 
contours of the 100$\mu$m emission. 

It is tempting to speculate on a possible link between the X-ray and FIR emission presented above, 
but we caution that this may not be real.
One possible factor affecting the morphology of X-ray emiting gas is the amount of foreground
HI, which modifies mainly the low energy end of the X-ray spectrum. Shadows in the X-ray background
caused by foreground HI clouds have been detected mainly in soft X-rays by \eg Egger 
\& Aschenbach (1995). However, the feature detected in the
ROSAT maps by B\"{o}hringer \etal (1994) is seen in the hard energy band (0.4--2.4 keV), and is
thus difficult to attribute it to gas absorption.

EUVE observations of the Virgo cluster (VC) center (Lieu \etal 1996) show the presence of gas
at $\sim$0.5 10$^6$ K near M87. This matter forms an additional component of the 
intra-cluster material (ICM) in Virgo, as follows from their analysis, and cannot be 
the same hot gas which is responsible for the
X-ray emission detected by ROSAT. In order to confirm the existence of this second ICM
component of the VC, Lieu \etal (1996) performed HI 21 cm observations with the 43-m Green
Bank telescope (angular resolution 21'). The region surveyed by them was centered on M87,
had an extent of 2$^{\circ}\times1^{\circ}.6$, and the grid of HI measurements was spaced 
every 8'$\simeq$1/3 of a resolution element. 

A comparison of the HI map of Lieu \etal (1996) with the FIR distributions (see
Figure 1) demonstrates that the FIR emission
follows the total HI column density. Although Lieu \etal  do not mention
the velocity range over which the Green Bank observations were performed, we assumed these to be
at $\sim$0 km s$^{-1}$ because they are supposedly of ``Galactic HI'' origin.
While some VC galaxies do have negative heliocentric velocities (c.f. Binggeli \etal 1985),
they mostly concentrate at 1,000--2,000 km s$^{-1}$. For this reason, we think it likely that
the HI detected by Lieu \etal (1996) does indeed belong to the MW and, by inference, so does the
material producing the FIR emission.
We note at this point that the center of the low N(HI) region, at 12:28+12:45 (1950) and only
$\sim$half a degree away from M87, has N(HI)$\simeq$1.8 10$^{20}$ atoms cm$^{-2}$. This is
coincident with the low FIR emission region. The ridges with the higher N(HI) values
correspond to enhanced FIR emission regions.

We produced N(HI) plots for the region using data from the Leiden-Dwingeloo
HI survey (Hartmann \& Burton 1997, LDS) in order to confirm the HI distribution
measured by Lieu \etal (1996).
The LDS was conducted with the 25-m radio telescope at Dwingeloo and the data we used
cover the velocity range --459$<v_{lsr}<$+415 km s$^{-1}$ with a resolution of 1.03 km s$^{-1}$. 
The 25-m radio telescope has
a 36' half-power beam and the survey was performed with 0$^{\circ}$.5
spacings. We used the file TOTAL\_HI.FIT from the
CD-ROM supplied with the printed atlas to extract the proper sky region.  The data
were transformed from Galactic to equatorial coordinates, accounting for the change
of scale from one side of the image to the other. This was done by dividing each 
pixel value by its {\it cos(b)}, to yield consistent units over the field.
The HI total column density from the LDS is shown in Figure 6 
together with the IRAS 100$\mu$m map and confirms the general
impression from the Lieu \etal (1996) map. The HI distribution
has a region with lower N(HI) at the center of the Virgo Ring (VR) and 
ridges of higher HI emission on both sides of the VR.

We also produced position-velocity (PV) plots using the channel data from the LDS, limiting these
to l=276.0, the galactic longitude of the HI peak which coincides with the main FIR emission
blob in area A of Fig. 6 (the more intense of the FIR peaks),  to the center of the VR (l=283.0),
and to the second highest FIR peak at l=290.5. The PV plots are shown in Figure 7 and indicate that 
at the position of the VR there is a significant disturbance of the HI, with a 
strong extension to negative velocities appearing in the PV plots of the high-FIR region. 
The sheet-like HI distribution, which links HI at low latitudes with gas near the Galactic 
Pole and has a slightly negative LSR velocity, appears disturbed at
b$\approx$75$^{\circ}$. 

The velocity plot through the peak  emission at l=276.0 at this latitude (Figure 8) shows three
peaks separated by $\sim$20 km s$^{-1}$. The strongest has the most negative 
velocity, approximately --30  km s$^{-1}$ and a FWHP of $\sim$11 km s$^{-1}$. The weakest peak 
at this location is near +4 km s$^{-1}$ (LSR). The PV in the low FIR region at the center of 
the VR (l=283$^{\circ}$, b$\approx$75$^{\circ}$) shows a single strong peak at $\sim$--7 km s$^{-1}$ 
(LSR), with a FWHP of 12.5 km s$^{-1}$ and a low shoulder extending to more negative velocities, 
down to the velocity of the strong peak at the location of the main blob (--30  km s$^{-1}$). The third PV,
at (l=290$^{\circ}$.5, b$\approx$75$^{\circ}$) is narrow  with a FWHP of $\sim$ 5km s$^{-1}$ and 
is centered at --7 km s$^{-1}$ (LSR).

\section{Discussion}

We identified a ring-like feature of FIR emission at high galactic latitude,
which is distant from the main body of the Galaxy and extinguishes light from galaxies
in the central part of the Virgo cluster (VC). There is no way to establish a distance to
the extinguishing cloud with the data we presented above, except to note that it is
probably $>$1 kpc. 

A nearby dust feature, observed by Haikala \etal (1995) in the
far-UV, has been located at $\sim$120 pc using the distribution of E$_{b-y}$ color
excesses. This dust cloudlet produces a visual
extinction A$_V\leq$0.4 mag and is located at (l=251.1, b=+73.3); this location is very 
similar to what we found for the Virgo Ring (VR) and may indicate that either our distance evaluation
is wrong, or that the location technique of the Haikala \etal feature did not use 
a sufficient number of more distant stars.

Indications that the dust cloud cannot be a nearby feature originate mainly from
the lack of influence on the distribution of stars. Supporting evidence to the same
comes from the reddening study of Knude (1996). He used uvbyH$\beta$ measurements of
A3-G0 stars with B$\leq$11.5 mag and $\vert$b$\vert>70^{\circ}$ to
determine the distribution of extinction. His results for E$_{b-y}$, broken by 
galactic latitude and by longitude quadrants, are of particular interest. The area of
interest for our study is located between the 3rd and 4th quadrants at b$\approx75^{\circ}$;
the reddening to this region is small, E$_{b-y}\leq0.017$,
which translates into A$_B\leq$0.095. The stars studied by Knude (1996) are
closer than 1.5 kpc (for main sequence A stars brighter than 11.5 mag), thus
the color-dependent extinction of the VC galaxies we detected, which is equivalent 
to A$_B\approx$0.4 mag, should be produced by material more distant than 1.5 kpc.

If the cloud would be in the VC itself, its physical
size would be $\sim$1.5 Mpc, very large indeed !
The issue of possible diffuse dust in clusters of galaxies has been studied by \eg 
Ferguson (1993). He concluded, from the lack of a difference between cluster and field galaxies
in the correlation of the Mg$_2$ index and (B--V), that dust is not present in the Virgo
cluster (upper limit E(B--V)$<$0.06 mag.). A similar conclusion for a large number of Abell
clusters, based on the (V--I) color indices of radio quasars seen in their background, 
was reached by Maoz (1995).

Not accounting for foreground dust may affect adversely some key
observations.
Our finding confirms the supposition of Zonn (1957) and Zonn \& Stodolkiewicz (1958),
that because of the patchy structure of the interstellar dust {\it it is not 
enough to correct for extinction assuming that the
dust is localized in a narrow slab near the Galactic equator, but the detailed 
distribution of dust must be investigated to account properly for the extinction.} In particular, 
many observations of the Virgo cluster and of objects within (\eg the HST Cepheid Key Project: 
Graham \etal 1998, Macri \etal 1999) may carry significant errors because of improper 
extinction corrections.

In this section we estimate the
%type and column density of 
dust temperature and dust-to-gas ratio.
%using the FIR emission properties, and compare these with the
%upper limits of dust obtained from the two optical studies described above.
To evaluate the temperature of the dust in VR we subtracted from the 
map intensities the minimal value for the central part of the VR, near M87, in 
the 100 and 240$\mu$m COBE/DIRBE bands (8 and 3.5 MJy sr$^{-1}$ respectively).
To determine the color temperature 
for the dust in the VR cloud we assumed that  the dust particles are in 
 thermal equilibrium and that the foreground galactic IR radiation and IR emission from all
point sources in the region have been subtracted accurately. These
assumptions may not be necessarily fulfilled in our case. The subtraction of the intensity
of the inner part of the ring would be accurate only if the distribution of the
foreground galactic radiation is fairly smooth; this is not the case even at high galactic latitudes.
Some galactic  features may add non-negligible FIR contributions to the foreground.
The radiation from very cold dust grains (T$\approx$3K) could not
be detected by the means used here. 
We could not rule out the possibility that transient heating
of dust grains takes place, and that only occasional excitation
by energetic photons or particles causes them to emit brief pulses of the radiation
measured by COBE/DIRBE and IRAS. 

With all these caveats in mind, we calculated the temperature
of two regions with maximal FIR intensities within regions delimited by: 
$\alpha$: 185$^{\circ}$--186$^{\circ}$, $\delta$: 13$^{\circ}$--14$^{\circ}$  (approximately the 
Main Blob=MB region) 
and $\alpha$: 189$^{\circ}$--190$^{\circ}$, $\delta$: 12$^{\circ}$--13$^{\circ}$ (slightly off the
secondary FIR peak). 
The temperature was calculated with the relation:
\begin{equation}
log_{10} T = 1.30274 + 0.26266(log_{10} R)+0.04935(log_{10} R)^2
\end{equation}
from Schlegel \etal (1998), where R=$\frac{I_{100}}{I_{240}}$ is the ratio of the
FIR intensities in the COBE/DIRBE bands. The results
are 22 and 20K for the first and the second region, respectively. These 
values can be accepted as upper limits for the dust temperature in VR due 
to both effects mentioned above. The temperatures do not differ significantly
from those usually adopted for interstellar dust clouds and are at the high end
of the range for the ``warm'' cirrus component (Lagache \etal 1998). The
difference between the two regions, with the MB being ``hotter'' than the secondary
FIR blob, is marginally significant, because the formal error in the derivation of
temperatures is 1-2K.

A study of the optical depth effects of high latitude clouds (Chlewicki \& Laureijs 1988)
found that typically $\frac{I_{100}}{N(HI)}$=0.7 MJy ster$^{-1}$/10$^{20}$ atoms cm$^2$.
Deul \& Burton (1990) studied the HI content and gas kinematics of  seven cirrus clouds detected
in the IRAS maps. They established that the FIR to HI ratio varies between
0.9 and 3.0 MJy ster$^{-1}$/10$^{20}$ atoms cm$^{-2}$. The color temperature
varies much less, between 0.20 and 0.27 for $\frac{I_{60}}{I_{100}}$.
Unfortunately, we did not detect 60$\mu$m emission from the VR, thus a direct comparison with the
color temperature derived by Deul \& Burton (1990) is not possible.
However, the ratio $\frac{I_{100}}{N(HI)}$ we find for the FIR ridges 
($\sim$1.5 MJy ster$^{-1}$/10$^{20}$ atoms cm$^{-2}$) is in the range of that measured 
for cirrus clouds. We determined N(HI) from the brightness temperature using data from
Fig. 9 and applying the conversion
\begin{equation}
N(HI)=1.8 \, 10^{18} \int{T_B(v) \, dv}
\end{equation}
%from Burke \& Graham-Smith (1997), 
assuming that the HI is optically thin.

The VR is located close to the northernmost point of the North Polar Spur (NPS) and
could, in principle, be part of it. The NPS is presumably
the remnant of a $\sim$15 Myr old explosive event in the direction of the Galactic center,
which released 10$^{56}$ ergs and could be
the outcome of $\sim10^5$ supernovae produced in a small region within one million years.
The giant loop is detected in a number of wavebands, from the X-ray
to the radio continuum (Sofue 1994). The possibility that the VR is part of a small 
structure, extending to lower galactic longitudes from near the northernmost apex 
of the large NPS feature in Sofue's Fig. 2 (Plate L11), cannot be excluded.

It is also clear that the Virgo Ring (VR) is not a known high velocity cloud, because the compilation
of Deul \& van Woerden (1990) lists as nearest object HVC 2 at 12:09+15:32 and v$_{lsr}$=100.7
km s$^{-1}$, covering 3.2 square degrees of the sky. 
Wakker \& van Woerden (1997) list a possible small HVC near the VR location. This 
is an HI feature associated with optical absorption lines observed in SN 1994D, 
located at l=290$^{\circ}$.07; b=+70$^{\circ}$.13 and $\sim$240 km s$^{-1}$ LSR. This 
SN occured in NGC 4526, and the absorbing gas (with five absorption systems)
was found by Ho \& Filippenko (1996) at the same velocity as the HI measured by Kumar \& Thonnard 
(1983). Ho \& Filippenko mention other similar absorption features in the spectrum of
SN 1991T (at l=292.59; b=65.18, in NGC 4527), which is $\sim5^{\circ}$ away from SN1994D.
It seems that there is material in the general 
direction of the VR with LSR velocities a few 100 km s$^{-1}$, and that it contains at 
least Ca II and Na I. In the vicinity of the VR, 
velocities in excess of 100 km s$^{-1}$ are required in order to qualify an HI feature as a HVC.

Our analysis of the HI distribution and kinematics, using the LDS survey data, indicates that
there is a significant disturbance of the hydrogen at the location of the VR. In particular,
the three-peak structure of the velocity-column density is peculiar. There are a number of possible
explanations for this HI peculiarity, starting from an expanding supernova (SN) shell, a
collision between Galactic HI and a high-velocity cloud, an ejection of a number of shells from a
red (super)giant, etc. Each of these possibilities imposes some constrains on the
problem.

Of all these possibilities, we consider the most likely to be that of a past collision between a small
HI cloud and Galactic hydrogen. The HI remnant of this event could be the feature seen
 at --30 km s$^{-1}$ in the high-FIR
area and the Galactic HI would be the prevalent 
emission at --7 to --9  km s$^{-1}$, which  shows up in all plots. At a velocity difference of
$\sim$20 km s$^{-1}$ the collision would be highly supersonic. We note that 
the immediate vicinity of the VR has been studied in exactly this
context by Stark \etal (1994). The region most likely to be part of the same complex is the BX
field (see Table 1 of Stark \etal 1994), where the X feature peaks near --8 km s$^{-1}$ (LSR) but 
there are parts of the B feature where velocities up to --40 km s$^{-1}$ are observed.

Stark \etal (1994) interpreted the ``intermediate negative velocity'' (INV) gas, with 
v$\leq$--20 km s$^{-1}$, as material at local velocity displaced by a shock following a
collision with a high velocity cloud. The difference in LSR velocity between the BX features
and candidate HVCs in this vicinity (at a few 100 km s$^{-1}$ positive) forced Stark
\etal to propose scenarios for the dissipation
of the impinging HVC. They asserted that this cloud has evaporated and is now
present as high-temperature gas. Collisions of HVCs with the Galactic disk have been studied by
Kerp \etal (1996). They show that a collision with a differential velocity of
25 km s$^{-1}$ may increase the temperature of the post-shock material from 10$^4$ 
to 10$^5$K; such temperatures are consistent with the EUVE measurements (Lieu \etal 
1996).

It is possible to estimate the parameters of the Main Blob (MB) using the HI information from the LDSS. 
With the T$_B$ value from --30 to --20 km s$^{-1}$
%, and following Burke \& Graham-Smith (1997),
we find
\begin{equation}
<N(HI)>\approx1.5 \, 10^{20} cm^{-2}
\end{equation}
This implies a gas-to-dust ratio for the MB of $\frac{N(HI)}{E(B-V)}\approx$5 10$^{20}$ cm$^{-2}$ 
mag$^{-1}$, different from the canonical value for the Galaxy 
($\frac{N(HI)}{E(B-V)}$=5.4
10$^{21}$ cm$^{-2}$ mag$^{-1}$; Bohlin 1975).
The MB appears, therefore, to be dust-rich (or HI-poor).
Its projected angular size is $\sim0^{\circ}.5$; this translates to a size
of  10($\frac{d}{1 \, kpc}$) pc and a total hydrogen mass of 
\begin{equation}
M_{tot}(HI)\approx100 \frac{d}{1 \, kpc} M_{\odot}
\end{equation}
with an average (volume) density of $\sim$4 $(\frac{d}{1 \, kpc})^{-1}$ cm$^{-3}$ assuming
a spherical configuration.
The cloudlet carries significant kinetic energy, considering its velocity relative to the Galactic HI
(assumed to be $\sim-9$ km s$^{-1}$); E$_K\approx$3 10$^{47} \, (\frac{d}{1 \, kpc})^2$ ergs.
Also, the HI profile observed in its direction (see top panel of Fig. 9) is
significantly wider than that of the second blob (bottom panel of Fig. 9). This could
be an indication of higher turbulence within the MB relative to other HI entities in
the same vicinty but at less negative LSR velocities.

It is significant that the MB, an infalling HI gas cloudlet, contains dust grains as
evidenced by the extinction it produces. This
indicates that any shocks that the material in MB might have encountered did not heat up its
material to temperatures high enough to completely destroy the grains. However, its highly supersonic
interaction with the ambient HI distribution caused a wider 21 cm profile.
The MB produces significant extinction and is fairly
distant from the MW plane while having an intermediate LSR velocity. This is 
different from the findings of Knude \& H$\o$g
(1999), who concluded that intermediate-velocity HI clouds show no extinction.

The detection of HI condensations on sub-degree scales and high above the Galactic 
plane is relevant for the studies of galaxies in the Virgo cluster and beyond. A spectroscopic survey 
of objects in the background clusters at z$\approx$0.08 could reveal absorption lines produced by
material within the MB cloudlet. This, along with a measure of the ionization presumably taking 
place at the interface between the MB and the ambient HI, could reveal its composition and location.
The essential finding is, however, the small scale on which significant extinction variations
are encountered at high $\vert$b$\vert$. This shows that the derivation of 
``average extinction dependences'', \eg with galactic latitude (Knude 1996), is not very 
relevant to the determination of cosmologically-important parameters from studies of individual 
VC galaxies.

Our study shows that dust can affect the light of background galaxies in the VC and, through this, the
 determination of H$_0$ by using
Cepheid photometry. This HST Key Project relies on a canned approach to deredden individual stars and
is based on the ``standard'' MW extinction law with R=$\frac{A_V}{E(B-V)}$=3.3 (c.f. Madore \&
Freedman 1991). It remains to be proven that this relation applies for high galactic latitude 
dust clouds, such as those studied here.

\section{Conclusions}

We detected a dusty HI cloud seen in projection against the central regions of the Virgo
cluster. We showed that the cloud, which appears ring-like in low-resolution FIR and
total HI maps, has a complex FIR morphology when examined with higher resolution.
The HI kinematics indicate a substantial disturbance at this location.

The Virgo Ring (VR) is located beyond the main body of the Galaxy, because the
  distributions and colors of stars in its direction are not affected by it. The cloud
 could not be more distant than the VC, because it influences the colors of galaxies
  within the cluster. The HI evidence argues strongly for a relatively nearby location.
The connection of
  the ring with the North Galactic Spur may be accidental, but the peculiar
morphology and kinematics of the HI associated with the FIR emission make a connection, perhaps through
a collision between an HI cloud and Galactic HI, more likely. Such a collision could
perhaps explain also the extreme-UV (or soft X-ray) emission observed from the neighborhood
of M87.

The importance of our finding lies in the possibility that many key observations
done in the direction of the Virgo cluster may have been adversely affected by
extinguishing clouds, which are not homogeneous and where the total V-band extinction
may reach up to a few tenths of a magnitude.

\section*{Acknowledgements}

NB acknowledges support from  the Israel Science Foundation. EA is
supported by a grant from the Israel Ministry of Science to develop TAUVEX, a UV imaging telescope.
KR and BW were supported by the Polish KBN grant
  number C76/98.
We are grateful to Drs. C.K. Young and M.J. Currie for supplying electronic versions of the VPC data
tables in advance of publication. NB is grateful to Mike Hauser for discussions 
on the reality of COBE/DIRBE
features, to John Bahcall for allowing public use of his Galaxy model,
and to Sara Beck and Sasha Kashlinsky for critical readings of one of the first drafts.

\section*{Appendix}

  Classical Wolf diagrams for galaxies (\eg Zonn \& Rudnicki 1965)  were made
for the regions labelled A through D in Fig. 4 and are shown in Figure 5. Only  curve C
could  be interpreted as shaped by extinction, and this only by a large amount of leeway. 
Curve A, as already cautioned, has a completely different shape. Both  A and D curves 
are higher than the
comparison curve B, and are presumably shaped by the inhomogeneous distribution of galaxies
in the VC area. If, however, we would adopt curve C as representative for the
background galaxies
(no VC subclusters or background galaxy clusters are evident in this area), then the
obscuring cloud would have to be relatively nearby, as no parts of the  C and B curves
overlap.

To prevent distortion of the results by possible subclustering, we 
decided to study the color distribution of the background galaxies.
    The advantage of using color is
    that this parameter is distance-independent, provided the objects of
    study are not ``too distant'', in the sense of requiring a k-correction.
    We assume that the objects in the VPC conform to this constraint and do
    not attempt to correct the color for redshift (which is not known, in
    most  cases). We also caution about ``edge effects'', which could arise
    because of a smaller population of the sub-areas located near the edges
    of the VPC coverage; such areas cover less than one square degree, have
less than the average number of galaxies, and
    would be less representative.

    We analyzed the color distribution of all galaxies in the entire VPC. 
We   selected objects only by their
    position; specifically, we included all the objects listed in the catalog
    with their magnitudes and colors, without separating them by
    morphological type, apparent blue magnitude, angular size, being members or
    non-members of the cluster, etc.
The area enclosed within 184$^{\circ}<\alpha<189^{\circ}$ and
    10$^{\circ}<\delta<16^{\circ}$ was divided into square degree areas at integer $\alpha$
    and $\delta$ values. In each sub-area we calculated the statistics of
    the color distribution for the galaxies. The total number of
galaxies per cell does not appear to show a specific pattern, apart for the
edge effects mentioned above.

We checked the distribution of photometric indices over the sub-areas of the VPC against the
locations of the enhanced FIR emission. In each one square degree area
we counted the galaxies, calculated their mean B$_J$ magnitude and the 
average (U--B$_J$), (B$_J$--R$_C$), and (U--R$_C$) color indices, and studied the
statistical properties of the distributions of the color indices in each area. The 
color index (U--R$_C$) has been calculated as 
(U-B$_J$)-(B$_J$-R$_C$). The results are presented in 
Tables A1 and A2 for (U--R$_C$), where each cell contains these parameters 
for each one square degree region.
The results for the other color indices were similar and are not shown.

 Each sub-area in Table A1 is represented by one cell in
    the table, where the rows are the declination of the sub-areas and the
    columns are the right ascension, both in decimal degrees.
    The distribution of the (U--R$_C$) color is represented by the mean, the
    standard error of the mean (in round brackets), and the number of
    galaxies in the sub-area [in square brackets]. The standard error of the
    mean $SE_a$ is defined as: 
    \begin{equation} 
    SE_a=\frac{SD_a}{\sqrt{N_a}} 
    \end{equation} 
    Here $SD_a$ is the standard deviation of the values in sub-area {\it
    a}, and $N_a$ is the number of galaxies in the same sub-area. 

    Table A2 shows the properties of the distribution of (U--R$_C$)
   in each of the one square degree areas. Specifically, we list
    the kurtosis with its standard error, and the skewness with its
    standard error, for the distribution of galaxies within the square
    degree area. These are listed in order to demonstrate how much are the
    distributions of (U--R$_C$) values in individual cells similar (or different).

The significance of possible differences among the cells was considered using the mean 
 and its standard error. A Student's {\it t-test} between the means, by which one
estimates the variable 
\begin{equation}
t=\frac{\vert<X_1>-<X_2>\vert}{\sqrt{SE_1^2+SE_2^2}}
\end{equation}
was performed. Here $<X_i>$ is the mean of variable {\it i}, and SE$_i$ is its standard 
error of the mean (\eg Brandt 1970).

A cursory perusal of Table A1 shows that the only sub-areas where (U--R$_C$) is small,
and which are not on the borders of the VPC (Currie \& Young 1998), are the two areas 
at ($\alpha$=187.5, $\delta$=13.5), which
we call ``area GA'', and at (187.5, 14.5) which we call ``area GB''. These regions are central 
in the Virgo Ring (VR) and this finding, if significant,
could indicate less wavelength-dependent extinction in this direction 
(which shows reduced FIR emission) than in neighboring areas. The
comparison is done with two regions located in the direction of enhanced FIR
emission, at (186.5, 12.5) which we call ``area GC'', and at (188.5, 12.5) which
we call ``area GD''. The latter area {\bf is} located
at the edge of the VPC but does not appear to be depleted in galaxies, thus the
comparison with it is valid.

Using the t-test described above, the difference in means of the
(U--R$_C$) color index between areas GA and  GC yields t=1.97 (significant at the 95\%
level), and between areas GA and  GD t=2.72 (signficant at the 99.5\% level). 
A similar significance level was found for the difference between areas GB and GC, with t=2.76.
The results of the t-test indicate that the galaxies in the VPC, which are located behind the low
FIR emission area, are slightly bluer than others which are located behind areas of high
FIR emission. This impression is supported by an examination of the higher moments of the 
distribution (Table A2). The skewness of the (U--R$_C$) values in areas GA and GB is very similar, 
but is very different from the skewness of the distribution in areas GC and GD.

\newpage

\section*{References}

\begin{description}

\item Abadi, H.J. \& Edmunds, M.G. 1978, A\&A, 70, 189

\item Almoznino, E. \&  Brosch, N.
1998, MNRAS, 298, 920

\item Bahcall, J.N. \& Soneira, R. 1984, ApJS, 55, 67

\item Beichman, C.A., Neugebauer, G., Habing, H.J., Clegg, P.E. \&
Chester, T.J. 1988 {\it IRAS Catalogs and Atlases: Vol. 1 (Explanatory
Supplement)}, NASA RP-1190

\item Binggeli,  B., Sandage, A. \& Tammann, G.A. 1985, AJ,  90, 1681

\item Blitz, L., Spergel, D.N., Teuben, P.J. \& Hartmann, D. 1999, astro-ph/9901307

\item Bohlin, R.C. 1975, ApJ, 200, 402

\item B\"{o}hringer, H., Nulsen, P.E.J., Braun, R. \& Fabian, A.C. 1995, MNRAS, 274, L67

\item B\"{o}hringer, H., Briel, U.G., Schwarz, A., Voges, V., Hartner, G. \& Trumper, 
J. 1994, Nature, 368, 828

\item Bothun, G., Impey, C.D., Malin, D.F. \& Mould, J.R. 1987, AJ,
    94, 23

\item Bothun, G., Schombert, J.M., Impey, C.D. \& Schneider, S.E.
    1990, ApJ, 360, 427

\item Brandt, S. 1970 {\it Statistical and Computational Methods in Data Analysis}, 
Amsterdam: North-Holland Publishing Co., pp. 125 {\it et seq.}

%\item Burke, B.F. \& Graham-Smith, F. 1997 {\it An Introduction to Radio
%Astronomy}, Cambridge (UK): The University Press

\item Currie, C.K. \& Young, M.J. 1998, A\&AS, 127, 367

\item de Blok, W.J.G. 1997, PhD thesis, University of Groningen

\item de Blok, W.J.G., van der Hulst, J.M. \& Bothun, G.D. 1995, MNRAS, 274, 235

\item Deul, F.R. \& Burton, W.B. 1990, A\&A, 230, 153

\item Egger, R.J. \& Aschenbach, B. 1995, A\&A, 294, L25

\item Ferguson, H.C. 1993, MNRAS, 263, 343

\item Ferguson, H.C., Tanvir, N.R. \& von Hippel, T. 1998, Nature, 391, 461

\item Giovanelli, R. \& Haynes, M. 1989, ApJ, 346, L5

\item Graham, J.A., Ferrarese, L., Freedman, W.L. \etal 1998, BAAS, 192, 66.12

\item Haikala, L.K., Mattila, K., Bowyer, S., Sasseen, T.P., Lampton, M. \& Knude, J.
1995, ApJ, 443, L33

\item Hartmann, D. \& Burton, W.B. 1997, {\it Atlas of Galactic Neutral Hydrogen},
Cambridge (UK): Cambridge University Press

\item Hauser, M.G., Kelsall, T., Leisawitz, D. \& Weiland, J. 1997, {\it COBE Diffuse Infrared
      Background Experiment (DIRBE) Explanatory Supplement}, COBE Reference
      Document Ref.No. 97A

\item Heller, A.,   Almoznino, E. \&   Brosch, N. 
 1998, MNRAS, in press

\item Ho, L.C. \& Filippenko, A.V. 1996, ApJ, 463, 818

\item Hoffmeister, C. 1962, Z.f.Astrophys., 55, 40

\item Kashlinsky, A. 1999, private communication

\item Katsiyannis, A.C., Kemp, S.N., Berry, D.S. \& Meaburn, J. 1998, A\&AS, 132, 387

\item Kerp, J., Mack, K.-H., Egger, R., Pietz, J., Zimmer, F., Mebold, U.,
Burton, W.B. \& Hartmann, D. 1996, A\&A, 312, 67

\item Knude, J. \& H$\o$g, E. 1999, A\&A, 341, 451

\item Knude, J. 1996, A\&A, 306, 108

\item Kumar, K.C. \& Thonnard, N. 1983, AJ, 88, 260

\item Lagache, G., Abergel, A., Boulanger, F. \& Puget, J.-L. 1998, A\&A, 333, 709

\item Leggett, S.K., Clowes, R.G., Kalafi, M., MacGillivray, H.T., Puxley, P.J., Savage, A.
\& Wolstencroft, R.D. 1987, MNRAS, 227, 563

\item Lieu, R., Mittaz, J.P.D., Bowyer, S., Lockman, F.J., Hwang, C.-Y. 
\& Schmitt, J.H.M.M. 1996, ApJL, 458, 5

\item Lo, K.Y. \& Sargent, W.L.W. 1979, ApJ, 227, 756

\item Macri, L.M. \etal 1999, astro-ph/9901332

\item Madore, B. \& Freedman, W.L. 1991, PASP, 103, 933

\item Maoz, D. 1995, ApJ, 455, L115

\item Murawski, W. 1983, Acta Cosmologica, 12, 7

\item Okroy, R. 1965, Astron. Cirk., 320, 4

\item Pickering, T.E., Impey, C.D., Van Gorkom, J.H. \& Bothun, G.D. 1997, AJ, 114, 1858

\item Rudnicki, K. 1986, {\it Gamov Cosmology} (F.Melchiorri \& R.Ruffini, eds.), LXXXVI Corso
                 of the Soc. It. Fisica, Bologna, p. 480

\item Rudnicki, K. \& Baranowska, M. 1966, Acta Astron., 16, 65

\item Schneider, S.E., Helou, G., Salpeter, E.E. \& Terzian, Y.
    1983, ApJ, 273, L1

\item Schlegel, D.J., Finkbeiner, D.P. \& Davis, M. 1998, ApJ, 500, 525

\item Sofue, Y. 1994, ApJ, 431, L91

\item Sprayberry, D., Impey, C.D., Irwin, M.J., McMahon, R.G. \&
    Bothun, G.D.  1993, ApJ, 417, 114

\item Stark, R., Dickey, J.M., Burton, W.B. \& Wennmacher, A. 1994, A\&A, 281, 199

\item Taylor, C.L., Thomas, D.L., Brinks, E. \& Skillman, E.D. 1996,
    ApJS, 107, 143

\item van den Bergh, S. 1996, PASP, 108, 1091

%\item van der Werf, P. 1989, PhD thesis, University of Groningen

\item Wakker, B.P \& Boulanger, F. 1986, A\&A, 170, 84

\item Wakker, B.P. \& van Woerden, H. 1991, A\&A, 250, 509

\item Wakker, B.P. \& van Woerden, H. 1997, ARAA, 35, 217

\item Weil, M.L., Bland-Hawthorn, J. \& Malin, D.F. 1997, ApJ, 490, 664

\item Wszolek, B., Rudnicki, K., de Bernardis, P., Masi, S. \& Salvi, A. 1988a, Ap.Space Sci.
                                                                         152,  29
  
\item Wszolek, B., Rudnicki, K., de Bernardis, P. \& Masi, S. 1988b, in {\it Large Scale Structures in the
                  Universe}, (Seitter, W.C., Duerbeck, H.W. \& Tacke, M., eds.) Lecture
		  Notes in Physics 310, 223

\item Wszolek, B., Rudnicki, K., de Bernardis, P. \& Masi, S. 1989, {\it The World of Galaxies}
             (Corvin, H.G., Jr. \& Bottinelli, C., eds.) p. 499

\item Zonn, W. 1957, Bull. Acad. Polonaise des Sciences, vol. V, no. 1, 47

\item Zonn, W. \& Stodolkiewicz, J. 1958, Bull. Acad. Polonaise des Sciences, ser. sci. 
math. astr. phys. vol. VI, no. 3, 185

\item Zonn, W. \& Rudnicki, K., 1965, in {\it Stellar Astronomy}, Washington DC, 168

\end{description}

\newpage

\section*{Figure captions}
\begin{description}
\item Figure 1: FIR maps of the VR region. The top right panel shows the
IRAS 100$\mu$m map. The COBE/DIRBE
100$\mu$m (top left panel) and 240$\mu$m emission (lower left
panel) show the VR, but have a lower angular resolution than the IRAS 100$\mu$m
 map. The contours of the FIR emission are in MJy ster$^{-1}$. For orientation purposes, the
location of M87 is (187.07, +12.67).

%\item Figure 2: IRAS maps of the same region as shown in Fig. 1, for the 60$\mu$m (right panel)
%and 100$\mu$m (left panel).

\item Figure 2: Wolf diagram for USNO-A2.0 stars in five selected regions, one corresponding 
to the lowest FIR-emitting region in Fig. 1 and the other four coinciding with peaks
of the FIR emission. The regions are defined in Table 1 and the cumulative
distributions are depicted as follows: USNO1222+1330 as a dotted line,
USNO1226+1430 as a dashed line, USNO1230+1130 as a long-dashed line, USNO1230+1330 as a
solid line with open squares, and USNO1234+1230 as a dot-dashed line.

\item Figure 3: Comparison of predicted star counts (using the Bahcall-Soneira model)
and the actual star counts from the USNO-A2.0 catalog, for the five regions listed in Table 
3. Each region is one degree square; the actual star counts are represented by open squares
and the model prediction is shown as the solid line.

\item Figure 4: Selected areas to check for total extinction effects on galaxies.
The contours refer to the IRAS 100$\mu$m map and the heavy-lined parallelograms marked A
through D are the selected areas for testing the number density of galaxies.

\item Figure 5: Wolf diagram for galaxies.  
The three distribution curves A, C, D correspond
to regions within higher FIR emission of Fig. 1 and curve B to the region of low
FIR emission. For explanation of the designations A, B, C, D see Fig. 5. The data from each of 
the regions is depicted as follows: region A - solid
line, region B - dotted line, region C - dashed line, and region D - long-dashed line.

\item Figure 6: Total HI column density from the LDS (left panel) compared
with the IRAS 100$\mu$m emission (right panel). Darker shades in the HI plot
correspond to higher column densities. Lighter shades in the FIR plot 
correspond to more intense emission.

\item Figure 7: Galactic latitude-velocity plots near b$\approx$75$^{\circ}$ for l=276.0 (top
panel), l=283.0 (middle panel), and l=290.5 (bottom panel). The first plot corresponds to the main
blob (more intense FIR emission), the second to the ``hole'', \ie center of the VR, and the third 
to the second blob (next intense FIR emission). The horizontal band marked with heavy lines 
in each panel indicates the region for which data were used to plot Fig. 8.

\item Figure 8: Cuts through the position-velocity plots of Fig. 7, averaging the brightness
temperature
 for b=74$^{\circ}, \, 74^{\circ}$.5, and b=75$^{\circ}$ (three latitude bands). The order of the plots
is like in Fig. 8. The average HI column density  
shows the velocity distribution of HI clouds in this region. Three peaks stand out clearly, with 
the most intense one at the highest negative velocities for the main blob.

\end{description}

\newpage
%  \begin{table*}[htb]
\begin{deluxetable}{cccccc}
\tablecaption{Regions for testing stellar properties}
\small
\tablehead{\colhead{Name} & \colhead{Center(1950)} & \colhead{Center(degrees)} & 
\colhead{(l, b)} & \colhead{N$_{stars}$}  }
\startdata
USNO1222+1330 & 12:22:00 +13:30:00 & 185.5 +13.5 & 277.3 +74.7 & 1804 \nl
USNO1226+1430 & 12:26:00 +14:30:00 & 186.5 +14.5 & 279.3 +76.0 & 1593 \nl
USNO1230+1130 & 12:30:00 +11:30:00 & 187.5 +11.5 & 286.4 +73.5 & 2126 \nl
USNO1230+1330 & 12:30:00 +13:30:00 & 187.5 +13.5 & 284.4 +75.4 & 1696 \nl
USNO1234+1230 & 12:34:00 +12:30:00 & 188.5 +12.5 & 289.0 +74.7 & 1669 \nl
\enddata

\tablecomments{Each region is a one degree square centered on the listed coordinates. The stars
were extracted automatically from the USNO-A2.0 catalog.}

\end{deluxetable}

\newpage
%  \begin{table*}[htb]
\begin{deluxetable}{ccccccc}
\tablecaption{A1: Photometric properties of VPC galaxies (U--R$_C$)}
\small
\tablehead{\colhead{$\delta$/$\alpha$} & \colhead{184.5} & \colhead{185.5} & 
\colhead{186.5} & \colhead{187.5} & \colhead{188.5} }
\startdata 

    15.5 & 1.485(.109)[08] & 1.300(.103)[04] & 1.360(.186)[06] &         
              1.678(.162)[08] & 1.751(.118)[12] \nl 
    14.5 & 1.347(.079)[27] & 1.476(.060)[48] & 1.547(.061)[46] &      
              1.380(.096)[26] & 1.526(.083)[30] \nl 
    13.5 & 1.573(.053)[68] & 1.508(.057)[56] & 1.545(.074)[39] &      
              1.390(.094)[26] & 1.445(.090)[22] \nl 
    12.5 & 1.363(.070)[36] & 1.602(.047)[57] & 1.608(.058)[51] &      
              1.593(.070)[45] & 1.693(.060)[35] \nl 
    11.5 & 1.274(.075)[36] & 1.582(.068)[32] & 1.658(.054)[72] &      
              1.514(.066)[45] & 1.420(.101)[26] \nl 
    10.5 & 1.296(.070)[11] & 1.347(.077)[22] & 1.530(.099)[23] &      
              1.596(.104)[20] & 1.054(.114)[14] \nl
 \hline

\enddata

\tablecomments{Each cell in the table contains the average (U--R$_C$) color
index, the standard error of the mean (in round brackets), and the number of galaxies in
area cell [in square brackets]. The cells, except those at the edges of the VPC coverage which are
smaller, are labeled by their central ($\alpha$, $\delta$)
coordinates, which are given in decimal degrees.}

\end{deluxetable}
\newpage
%  \begin{table*}[htb]
\begin{deluxetable}{ccccccc}
\tablecaption{A3: Statistical distribution properties of (U--R$_C$) in VPC galaxies}
\small
\tablehead{\colhead{$\delta$/$\alpha$} & \colhead{184.5} & \colhead{185.5} & 
\colhead{186.5} & \colhead{187.5} & \colhead{188.5} }
\startdata

15.5 & -.37(1.48) .11(.75) & 1.89(2.62) 1.35(1.01) & .83(1.74) .16(.85)
    & .74(1.48) -.46(.75) & .18(1.23) -.64(.64) \nl 
    14.5 & -.31(.87) .58(.45) & -1.38(.67) -.05(.34) & -1.10(.69) -.36(.35)
    & -.96(.89) .11(.46) & .35(.83) .00(.43) \nl 
    13.5 & .83(.57) -.86(.29) & -.32(.63) -.07(.32) & -.03(.74) .45(.38) &
    -.09(.89) .53(.47) & 1.89(.95) -.91(.49) \nl 
    12.5 & -.43(.77) -.33(.39) & 1.38(.62) -.71(.32) & .43(.67) -.47(.33) &
    .35(.70) -.64(.35) & -.93(.78) -.43(.40) \nl 
    11.5 & -.47(.77) -.08(.39) & -1.06(.81) .10(.41) & -.21(.56) -.10(.28) &
    -.10(.70) .38(.35) & -.86(.89) -.02(.46) \nl 
    10.5 & .59(1.28) -.56(.66) & -1.04(.95) -.55(.49) & -.58(.94) -.55(.48)
    & .14(.99) .57(.51) & -.41(1.15) .11(.60) \nl
\enddata

\tablecomments{Each cell contains the kurtosis and its standard error,
and the skewness and its standard error, for the distribution of the (U--R$_C$) color indices
in each cell area. The errors for each statistical estimator are given in round brackets.}

\end{deluxetable}

%IRAS 100mu contours=virg1.ps
%COBE 240mu=virgo10.ps
%COBE 100 mu=virgo8.ps
%IRAS 100 mu grey=vir0.ps
%IRAS 60 micron= no60.ps

\newpage
 
  %side-by-side from Dave Zurek
\begin{figure}
\vspace{15cm}
%\special{psfile=SRFh.ps angle=0 vscale=100 hscale=100 voffset=-100 hoffset=-100}
%\putplot{vir0.ps}{1in}{-90}{50}{50}{-350}{500}
\putplot{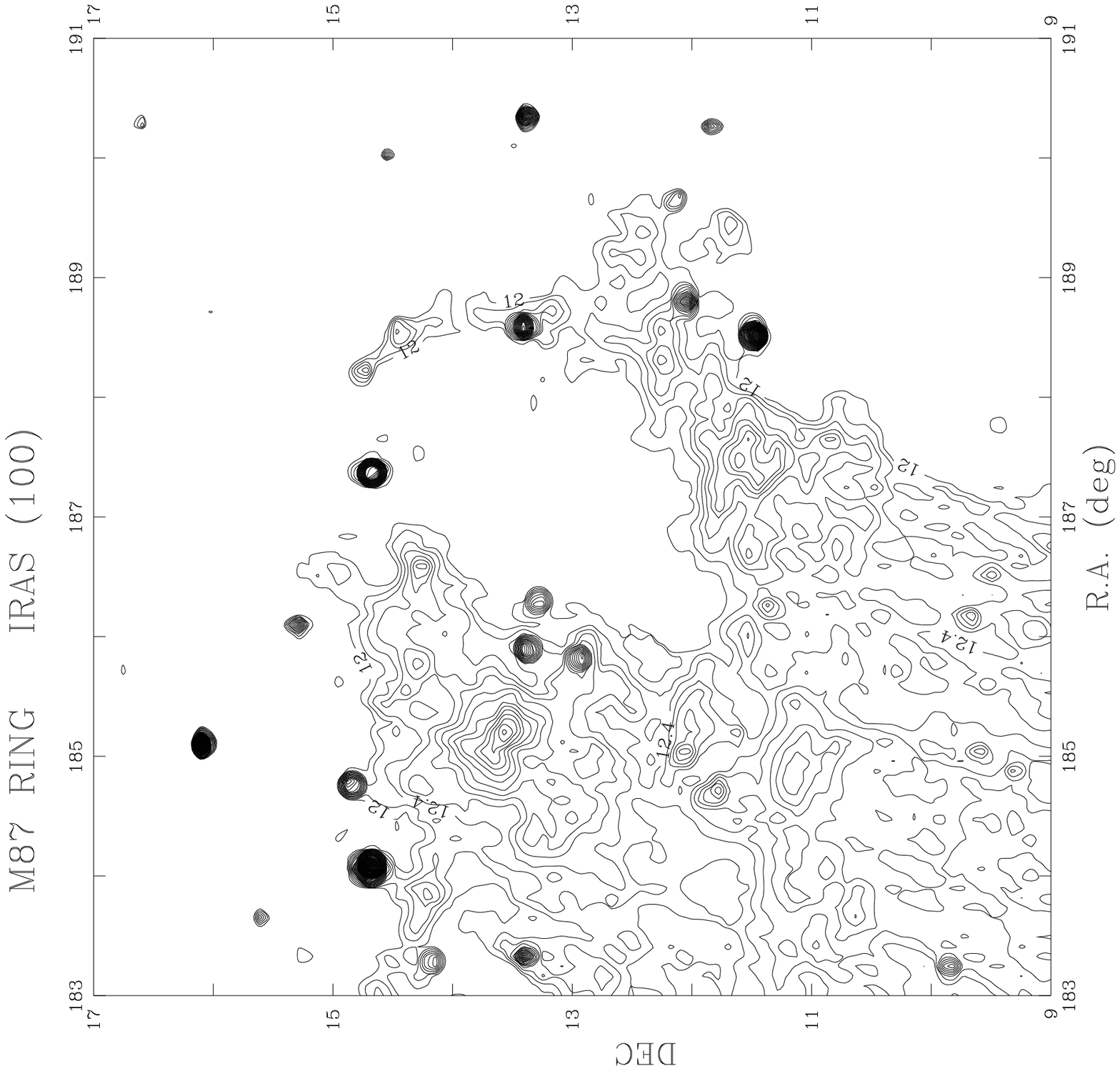}{1in}{-90}{50}{50}{-30}{560}
\putplot{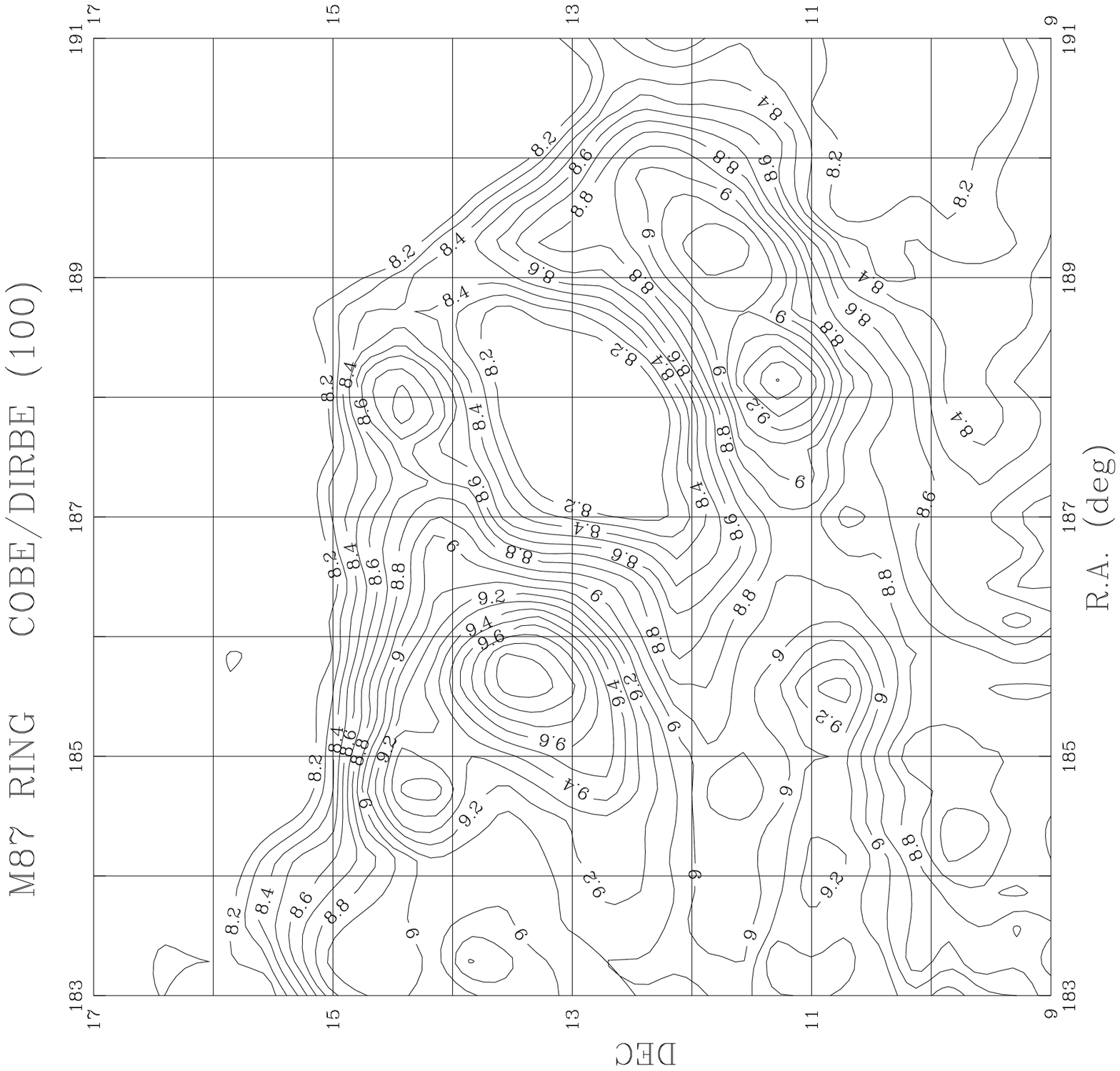}{1in}{-90}{50}{50}{-300}{650}
\putplot{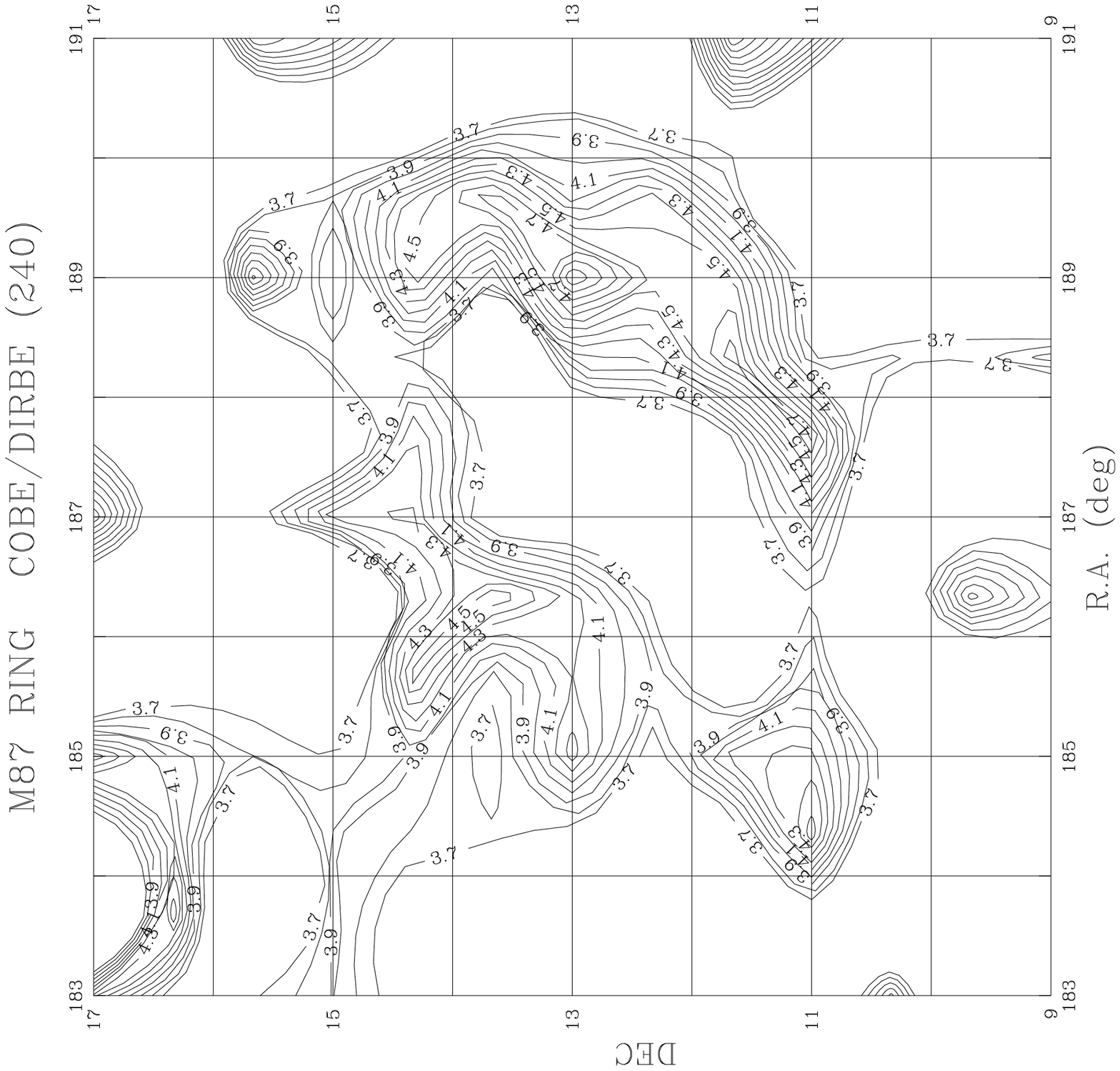}{1in}{-90}{50}{50}{-300}{450}
%\caption{FIR emission plots}
\end{figure}
\pagebreak

\newpage
 %Wolfall.eps overplot of all WDs:
\begin{figure}[tbh]
\vspace{14cm}
%\ref{Virgopred}
\includegraphics{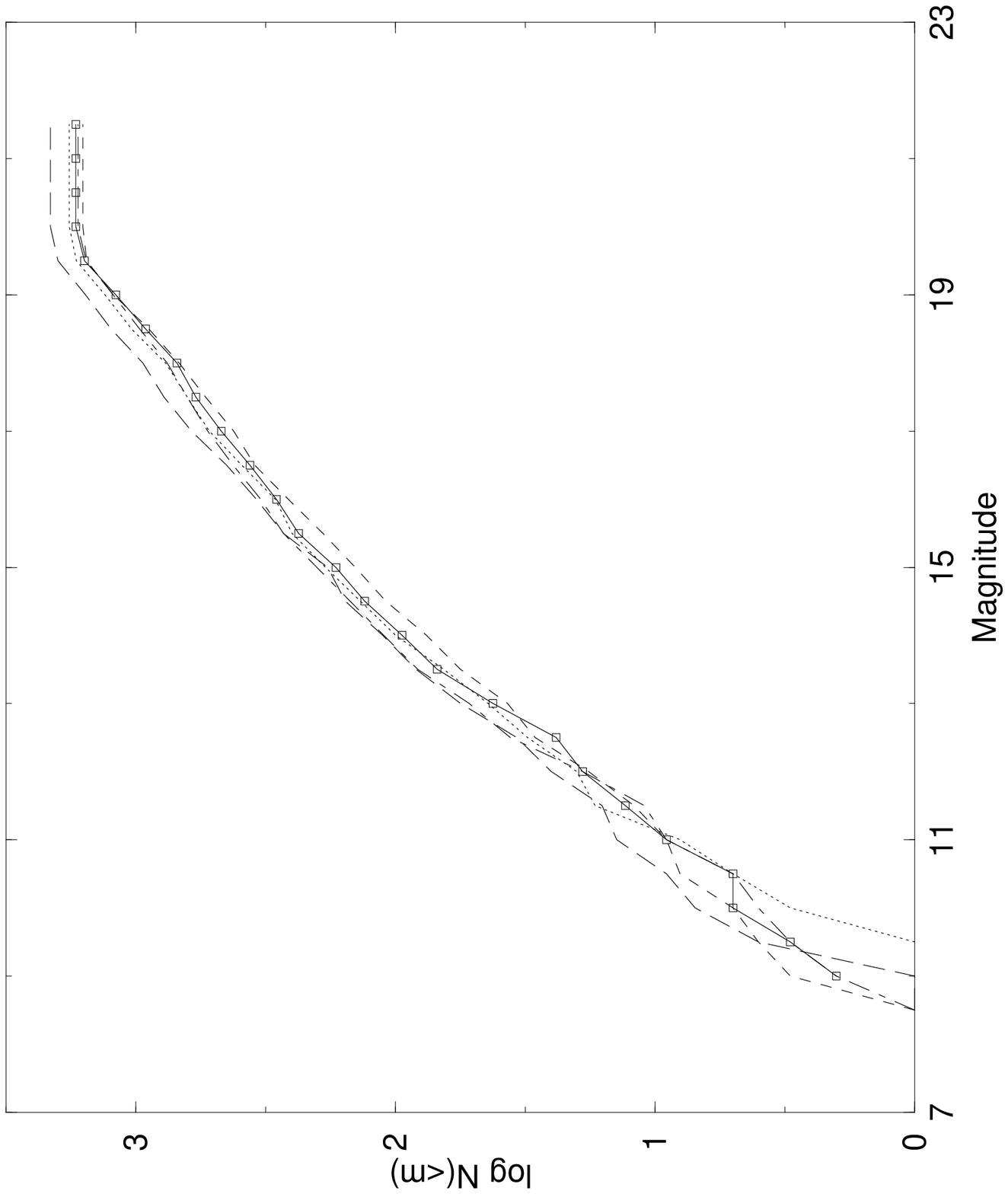}
%\caption{Single {\it vs.} multiple measurements of the same object. The
%solid line represents the distribution of FUV magnitudes for the objects observed
%more than once.}
\end{figure}

\newpage
  %side-by-side from Dave Zurek
\begin{figure}
\vspace{15cm}
%\special{psfile=SRFh.ps angle=0 vscale=100 hscale=100 voffset=-100 hoffset=-100}
\putplot{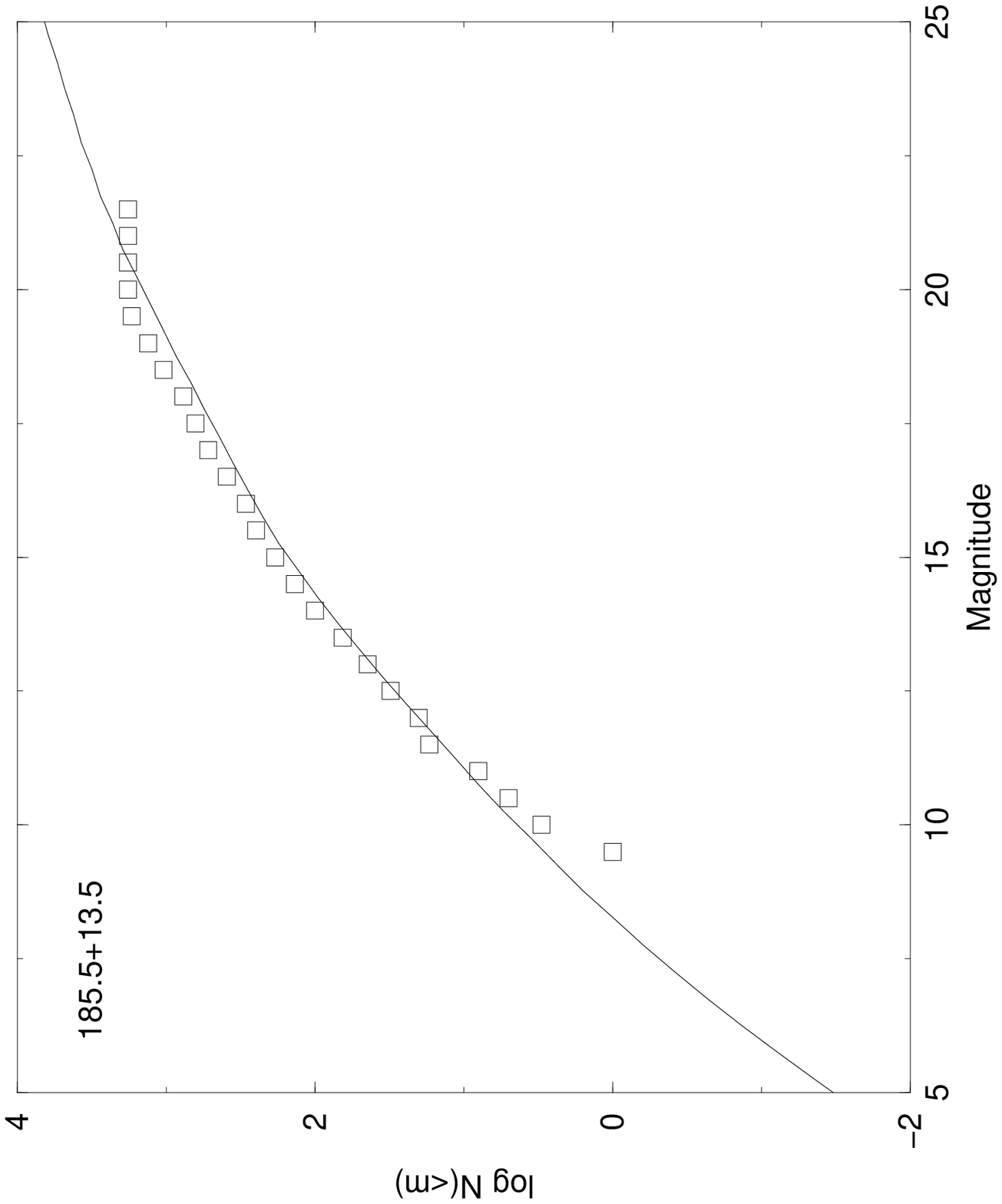}{1in}{-90}{50}{50}{-320}{555}
\putplot{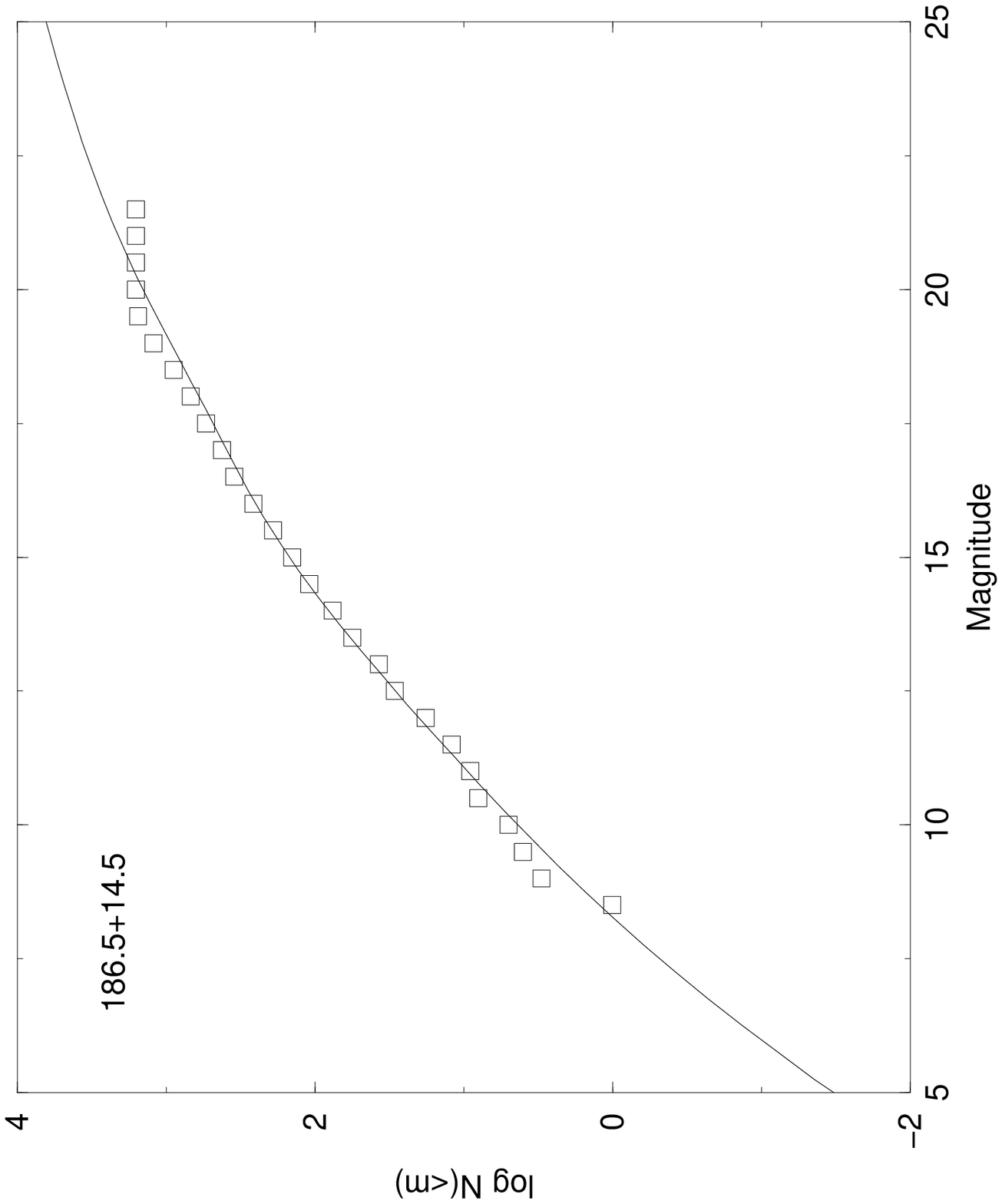}{1in}{-90}{50}{50}{-30}{650}
\putplot{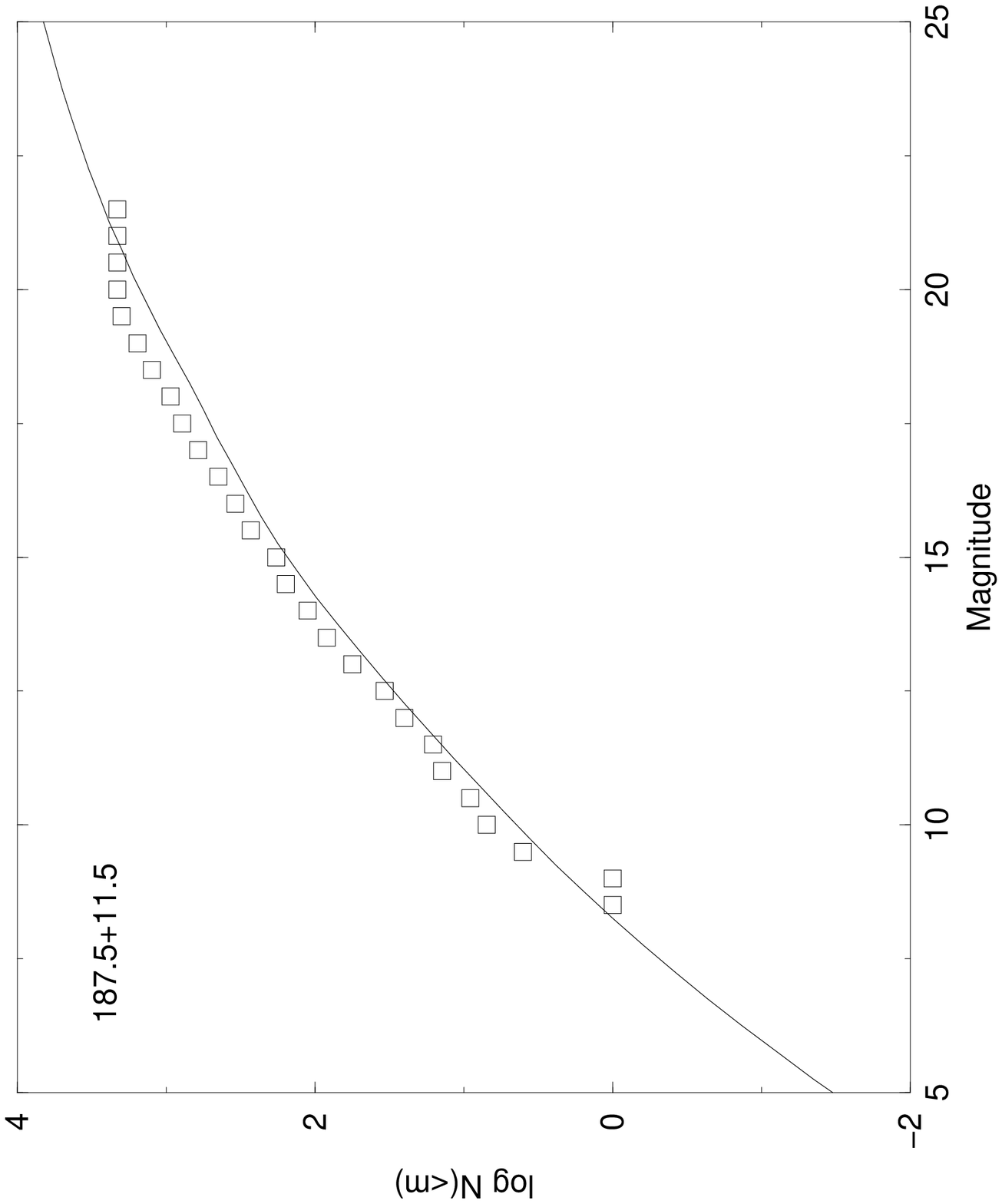}{1in}{-90}{50}{50}{-320}{500}
\putplot{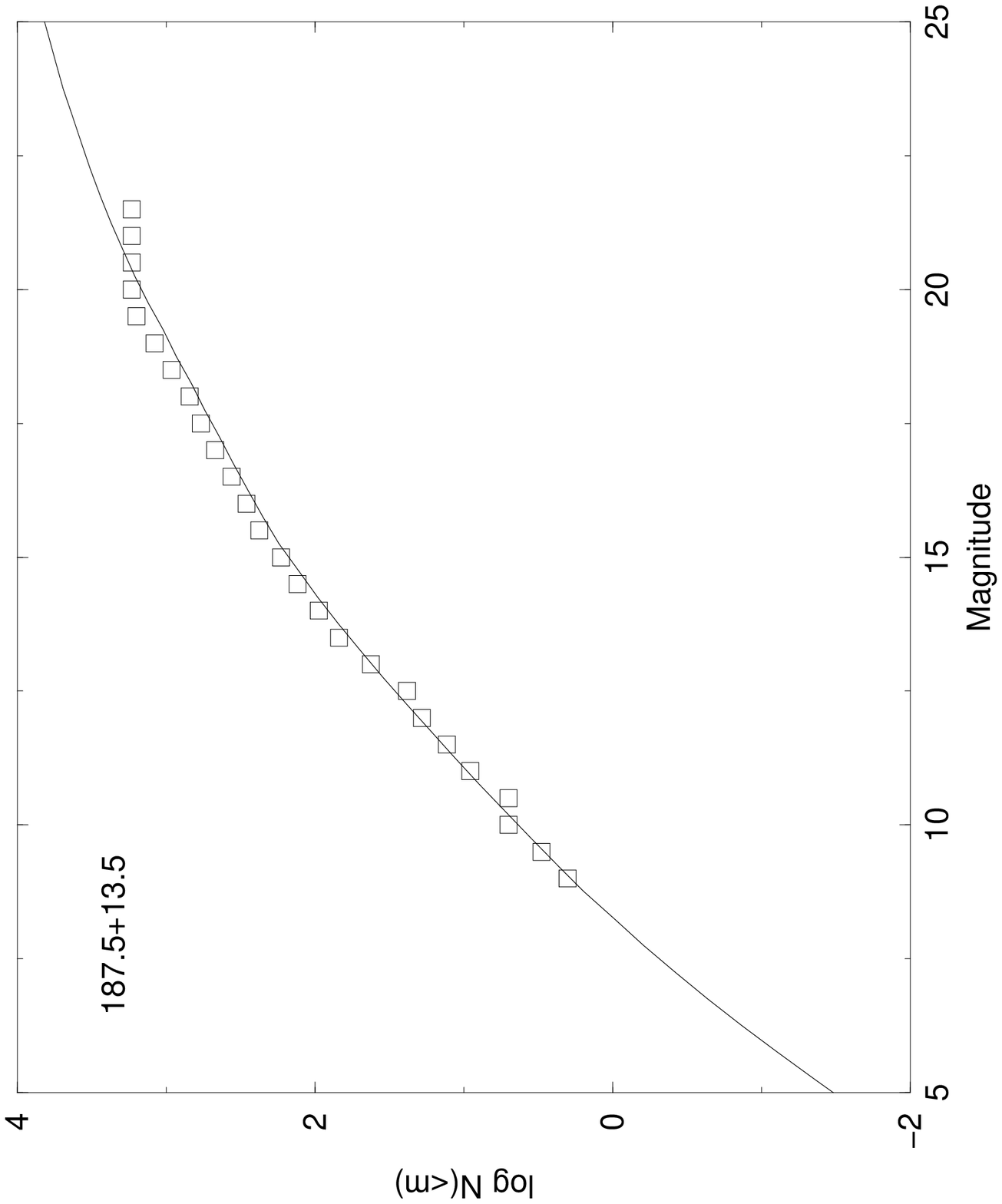}{1in}{-90}{50}{50}{-30}{600}
\putplot{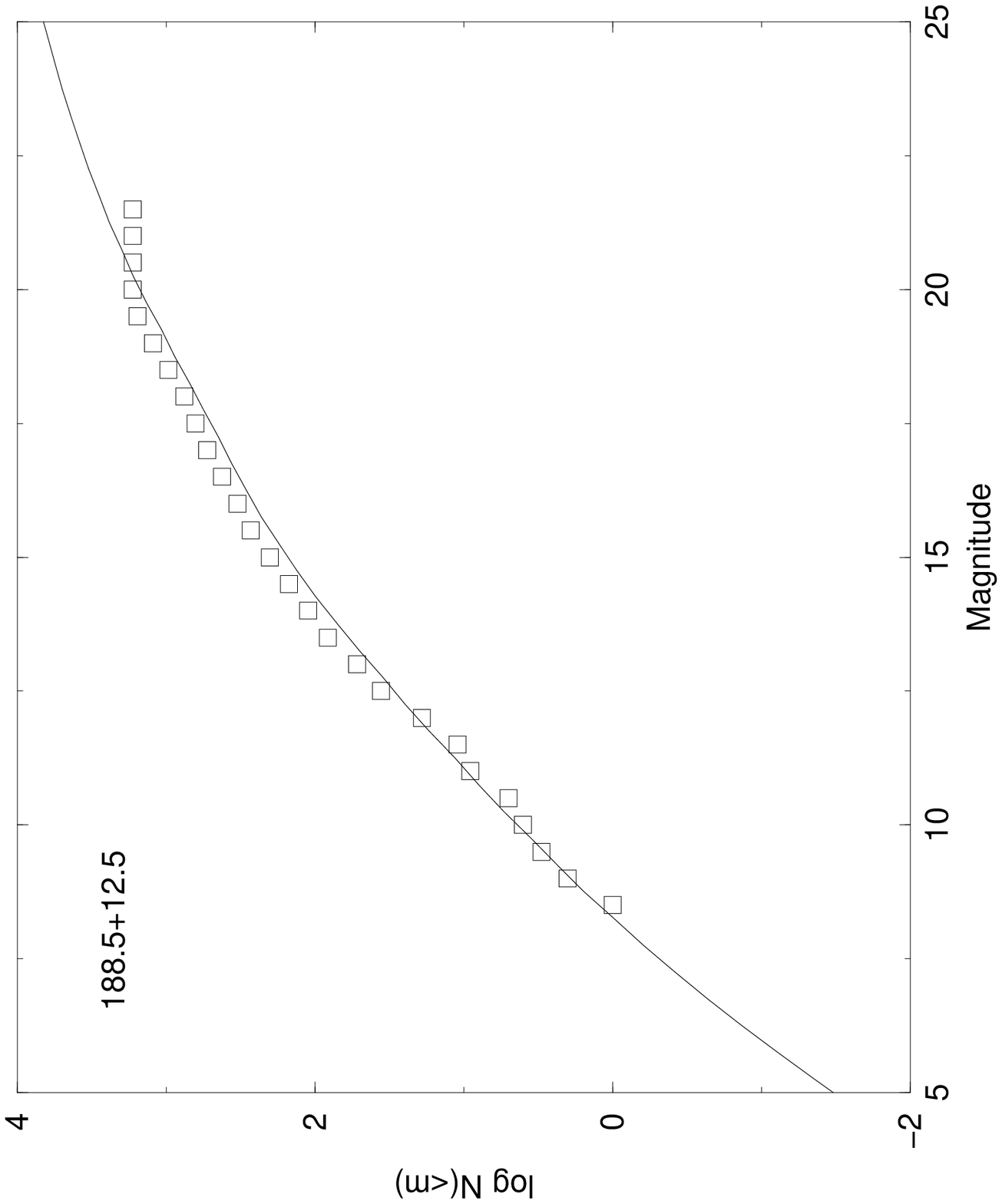}{1in}{-90}{50}{50}{-180}{450}
%\caption{Comparison of Wolf diagrams for stars, in various regions }
\end{figure}
\pagebreak
 \newpage

 %Wolfall.eps overplot of all WDs:
\begin{figure}[tbh]
\vspace{14cm}
%\ref{Virgopred}
\includegraphics{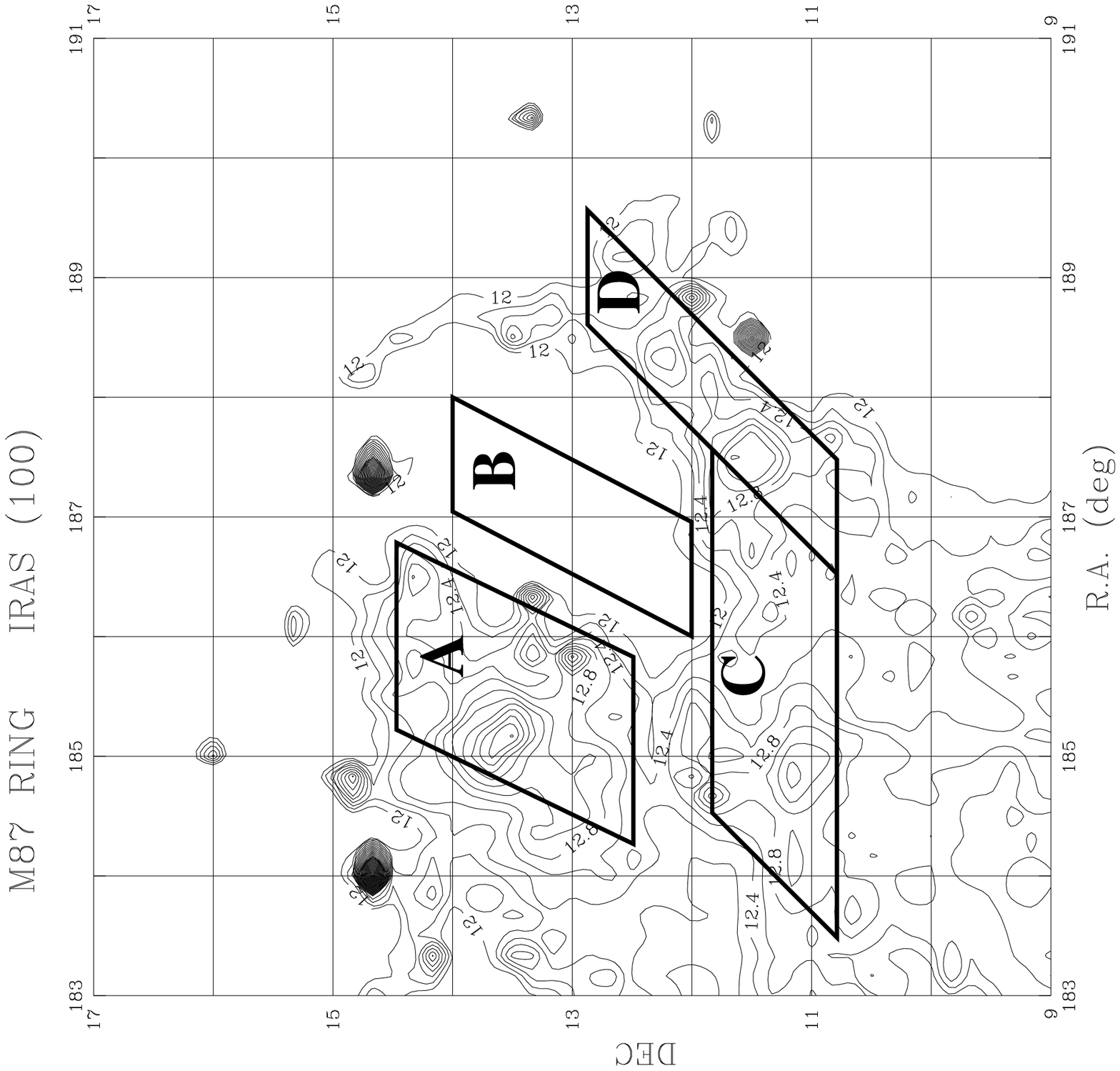}
%\caption{Overplot of areas for galaxies, for Wolf diag for galaxies.}
\end{figure}

\newpage
 %Wolfall.eps overplot of all WDs:
\begin{figure}[tbh]
\vspace{14cm}
%\ref{Virgopred}
\includegraphics{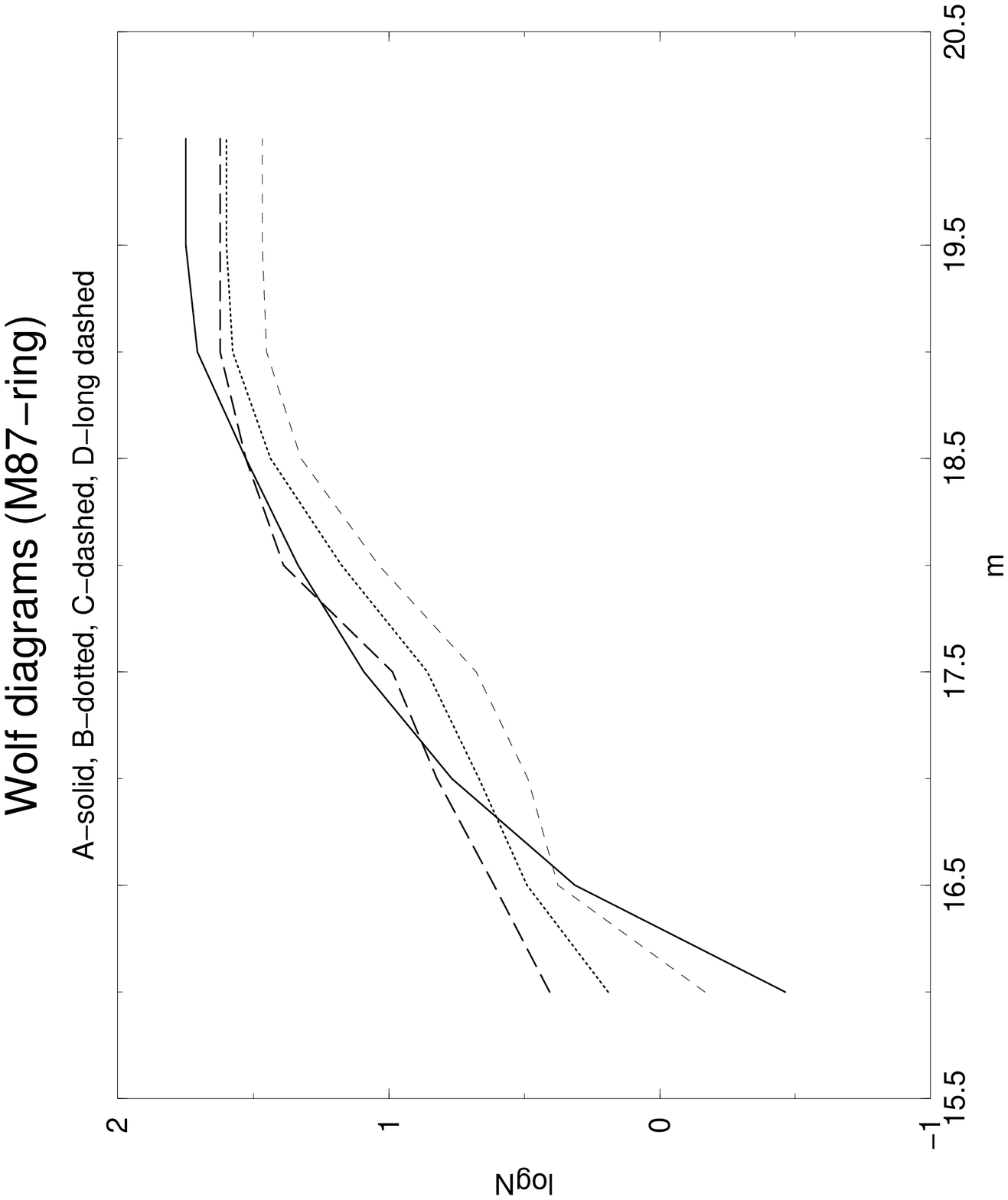}
%\caption{Wolf diagram for galaxies.}
\end{figure}

\newpage

  %side-by-side from Dave Zurek

\begin{figure}
\vspace{15cm}
%\special{psfile=SRFh.ps angle=0 vscale=100 hscale=100 voffset=-100 hoffset=-100}
\putplot{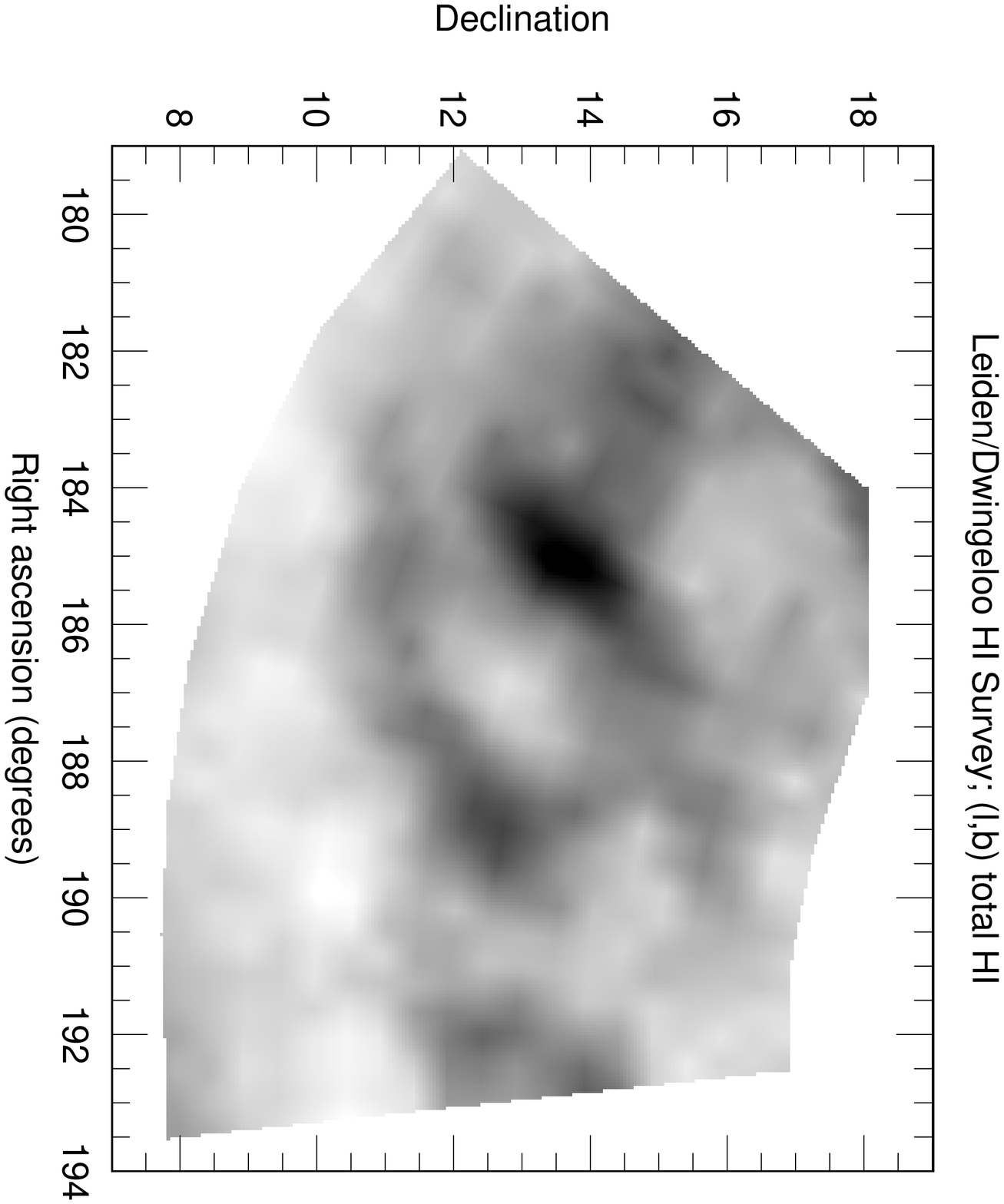}{1in}{0}{50}{50}{-310}{170}
\putplot{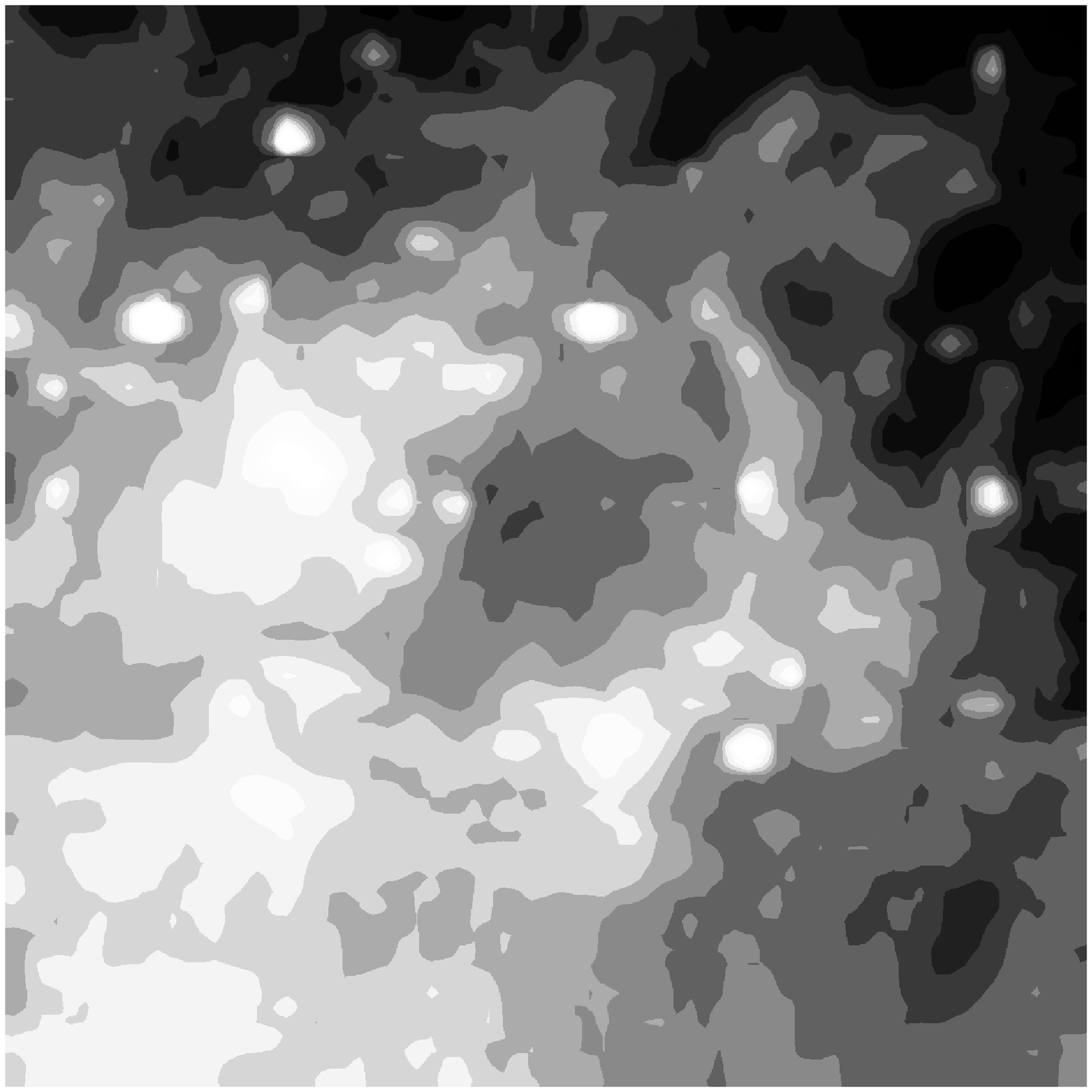}{1in}{90}{-40}{40}{-30}{310}
%\caption{Comparison of total HI column density distribution with the IRAS 100 micron map }
\end{figure}
\pagebreak
 \newpage

  %side-by-side from Dave Zurek

\begin{figure}
\vspace{15cm}
%\special{psfile=SRFh.ps angle=0 vscale=100 hscale=100 voffset=-100 hoffset=-100}
\putplot{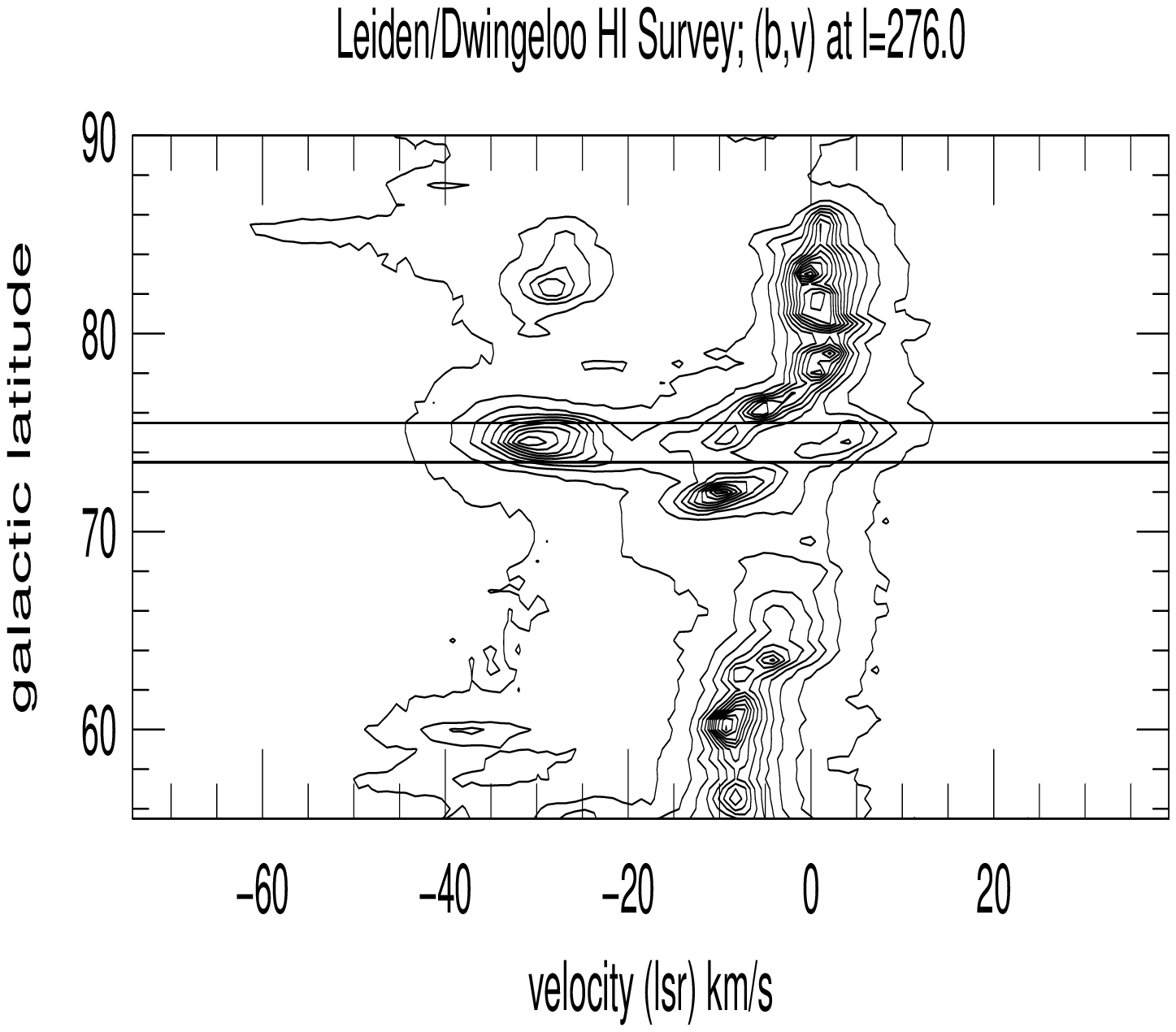}{1in}{0}{50}{50}{-150}{185}
\putplot{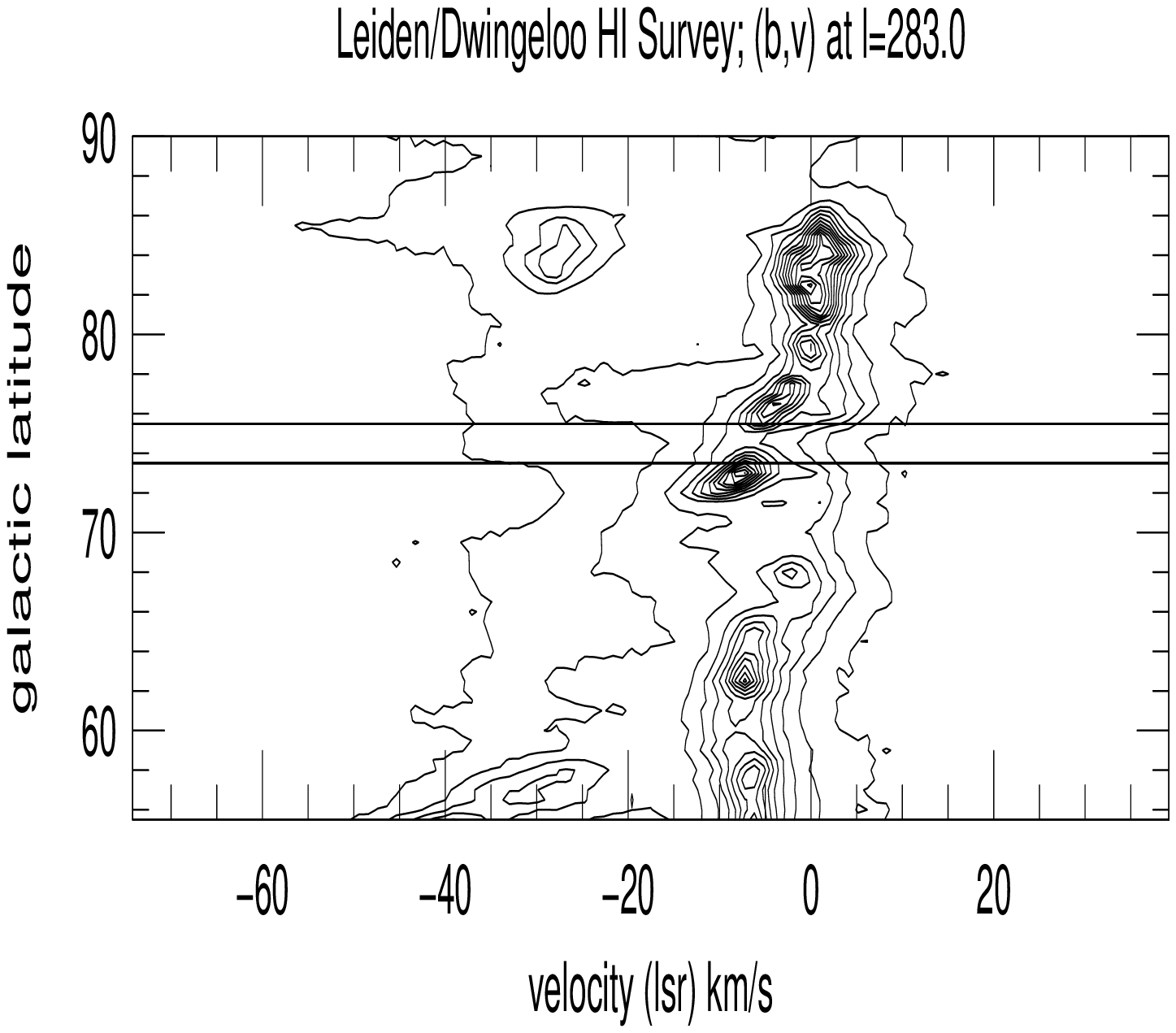}{1in}{0}{50}{50}{-150}{75}
\putplot{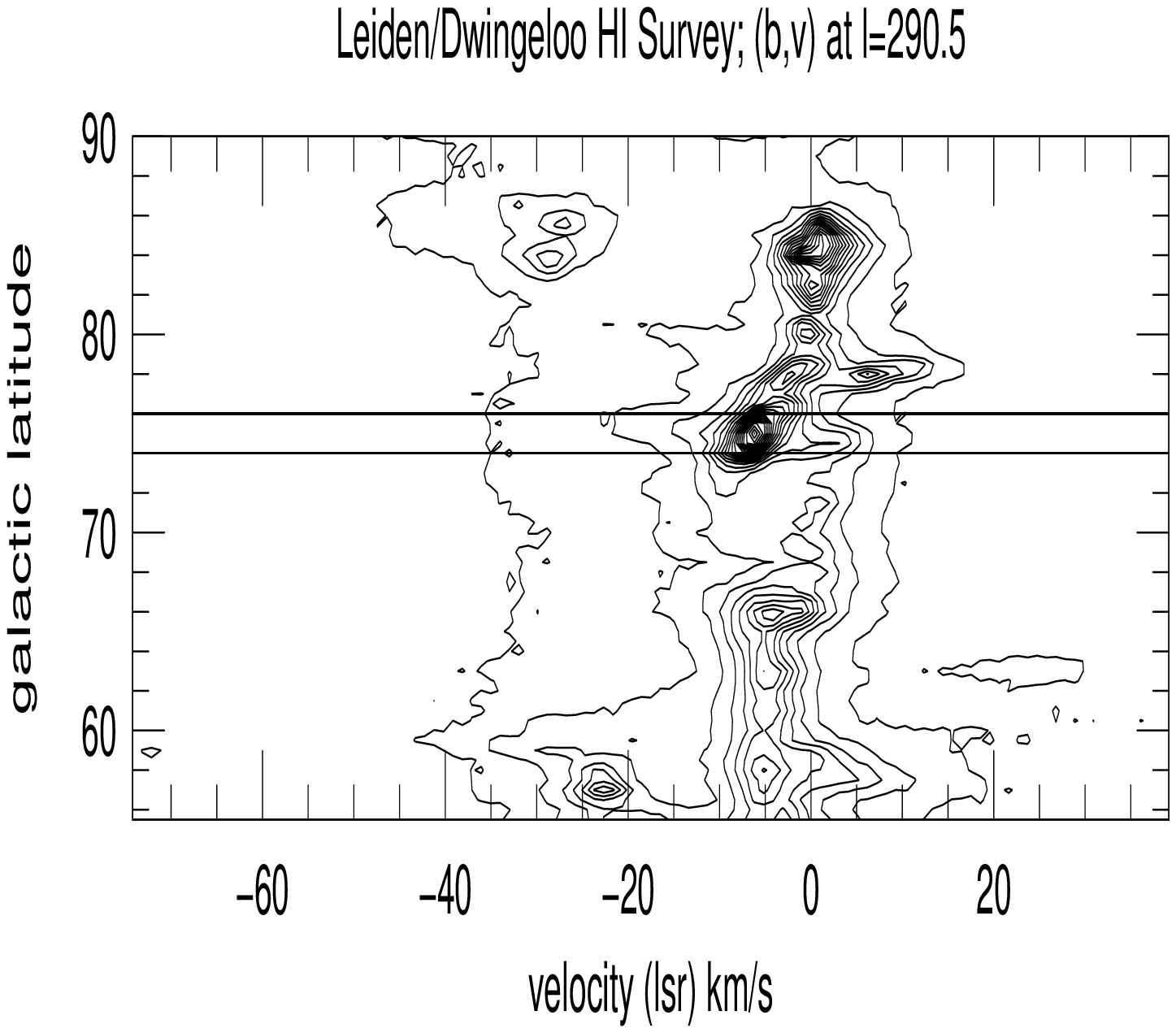}{1in}{0}{50}{50}{-150}{-35}
%\caption{Position-velocity plots for different longitudes}
\end{figure}
\pagebreak
 \newpage

  %side-by-side from Dave Zurek

\begin{figure}
\vspace{16cm}
\putplot {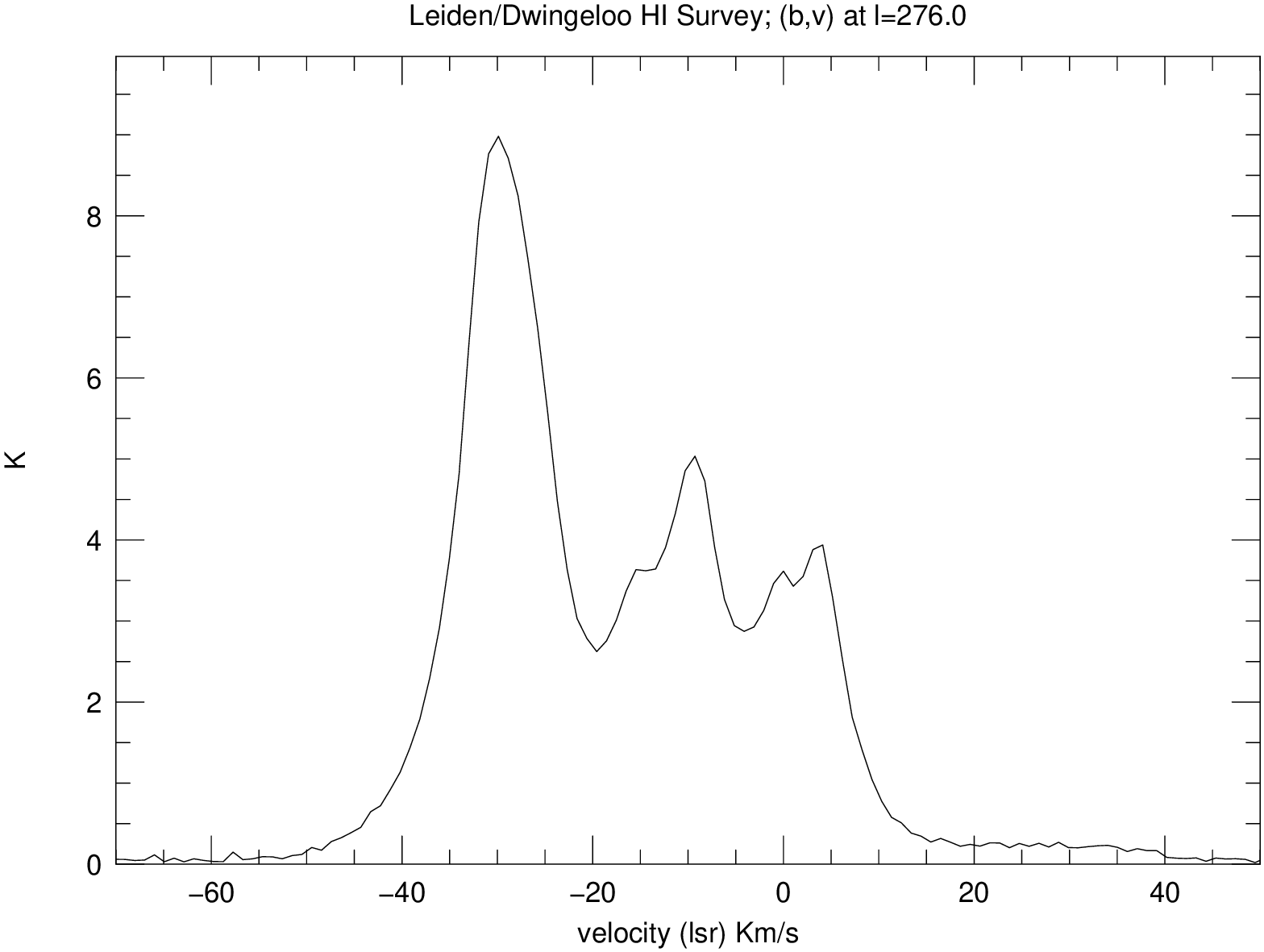}{1in}{0}{35}{35}{-250}{300}
\putplot {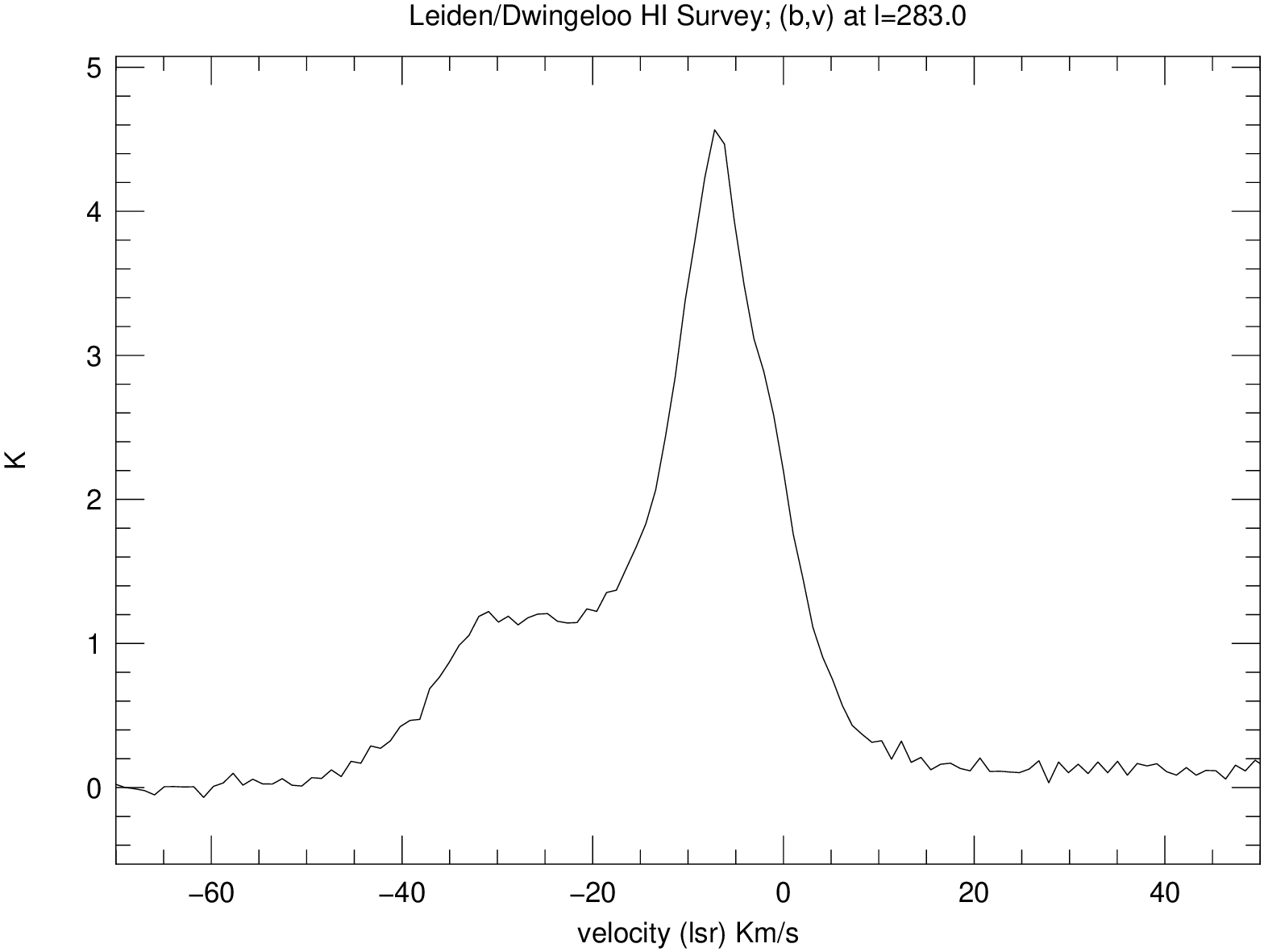}{1in}{0}{35}{35}{-250}{210}
\putplot{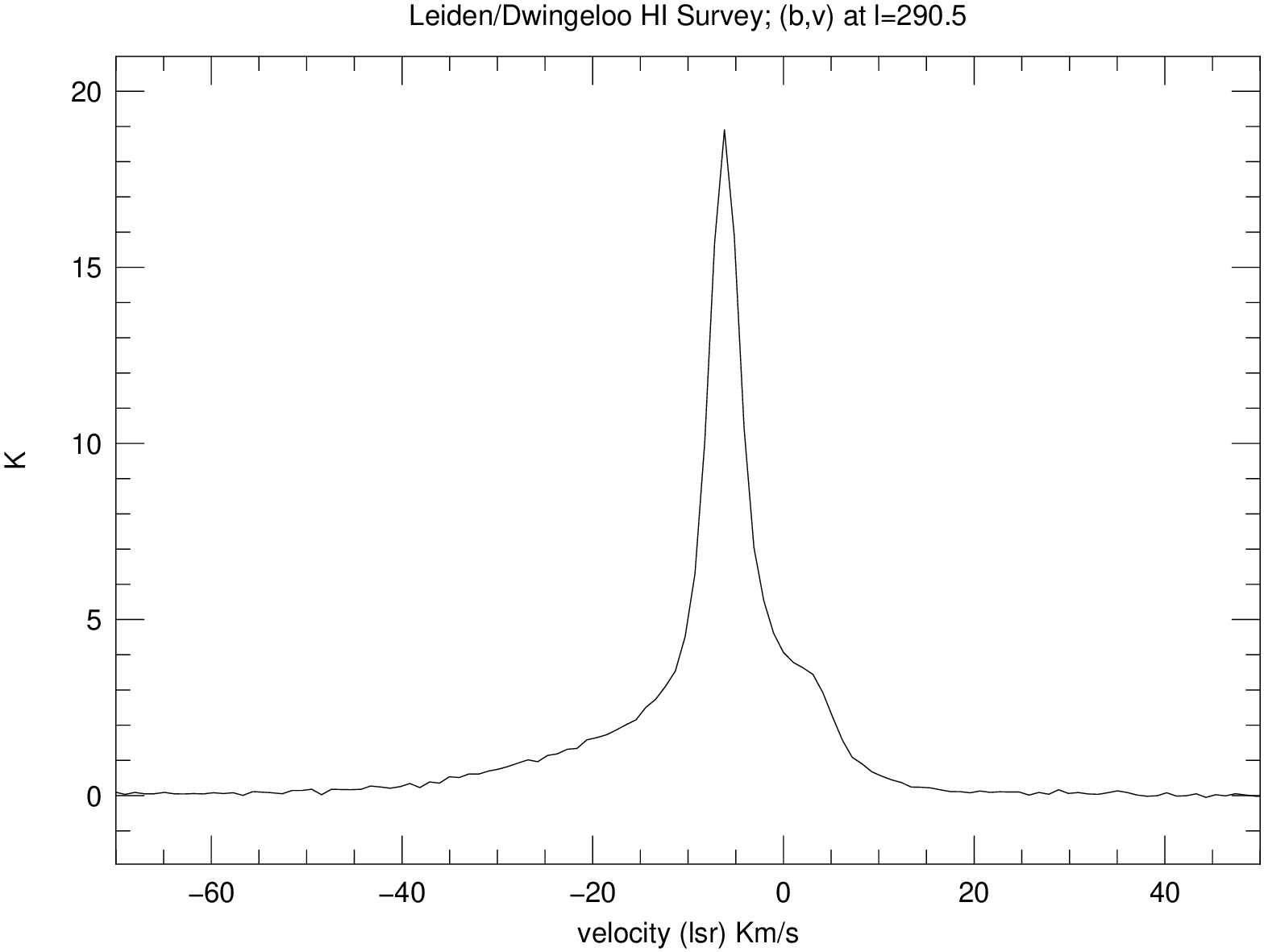}{1in}{0}{35}{35}{-250}{120}
%\caption{HI spectra at same locations as previous; new version from Nani}
\end{figure}
\pagebreak
 \newpage

\end{document}